\documentclass[useAMS,usenatbib,uselongtable]{mn2e}

\usepackage{lscape}

\usepackage{times}
\usepackage{graphicx}
\usepackage{xspace}
\usepackage{epsfig}
\usepackage{natbib}
\usepackage{rotating}
\usepackage{dcolumn}

\title[]{The UV and X-ray activity of the M dwarfs within $10$\,pc of the Sun}
\author[]{B. Stelzer$^{1}$\thanks{E-mail:stelzer@astropa.inaf.it} and A. Marino$^{2}$ and G. Micela$^{1}$ and J. L\'opez-Santiago$^{3}$ and C. Liefke$^{4}$ \\
$^{1}$ INAF - Osservatorio Astronomico di Palermo, Piazza del Parlamento 1,
 90134 Palermo, Italy \\ 
$^{2}$ Dipartimento di Fisica e Astronomia, G. Galilei, 
vicolo dell'Osservatorio 3, 35122 Padova, Italy \\
$^{3}$ Departamento de Astrof\'isica y Ciencias de la Atm\'osfera, 
Facultad de Ciencias F\'isicas, Universidad Complutense de Madrid, 28040 
Madrid, Spain \\
$^{4}$ Zentrum f\"ur Astronomie der Universit\"at Heidelberg, 
M\"onchhofstrasse 12-14, 69120 Heidelberg, Germany 
}

\begin{document}

\date{Accepted 2013 February 4.  Received 2013 January 31; in original form 2012 November 15}
%Accepted 1988 December 15. Received 1988 December 14; in original form 1988 October 11}

\pagerange{\pageref{firstpage}--\pageref{lastpage}} \pubyear{2013}

\maketitle

\label{firstpage}

\begin{abstract}
M dwarfs are the most numerous stars in the Galaxy. They are characterized by strong
magnetic activity. The ensuing high-energy emission is crucial for the evolution of
their planets and the eventual presence of life on them.
We systematically study the X-ray and ultraviolet emission of a subsample of 
M dwarfs from a recent proper-motion survey, selecting all M dwarfs within $10$\,pc to
obtain a nearly volume-limited sample ($\sim 90$\,\% completeness). 
Archival ROSAT, XMM-Newton and GALEX data 
are combined with published spectroscopic studies of H$\alpha$ emission and rotation to 
obtain a broad picture of stellar activity on M dwarfs. We make use of synthetic model
spectra to determine the relative contributions of photospheric and chromospheric emission
to the ultraviolet flux. We also analyse the same diagnostics for a comparison sample of  
young M dwarfs in the TW\,Hya association ($\sim 10$\,Myrs). 
We find that generally the emission in the GALEX bands is dominated by the chromosphere
but the photospheric component is not negligible in early-M field dwarfs. The surface 
fluxes for the H$\alpha$, near-ultraviolet, far-ultraviolet 
and X-ray emission are connected via a power law dependence. We present here for the first 
time such flux-flux relations involving broad-band ultraviolet emission for M dwarfs. 
Activity indices are defined as flux ratio between the activity diagnostic and the bolometric 
flux of the star in analogy to the 
Ca\,{\sc II} $R^{\prime}_{\rm HK}$ index. 
For given spectral type these indices display a spread of $2-3$ dex which is largest for 
M4 stars. Strikingly, at mid-M spectral types the spread of rotation rates is also at its 
highest level. 
The mean activity index for fast rotators, likely representing the saturation level, 
decreases from X-rays over the FUV to the NUV band and H$\alpha$, 
i.e. the fractional radiation output increases with atmospheric height. The comparison to 
the ultraviolet and X-ray properties of TW\,Hya
members shows a drop of nearly three orders of magnitude for the luminosity in these bands
between $\sim 10$\,Myr and few Gyrs age. A few young field dwarfs ($< 1$\,Gyr) 
in the $10$\,pc sample bridge the gap indicating that
the drop in magnetic activity with age is a continuous process. The slope of the age decay
is steeper for the X-ray than for the UV luminosity. 
\end{abstract}


\section{Introduction}\label{sect:intro}

Research on M dwarfs has become of central interest for astronomy, partly motivated by 
their importance as hosts of habitable planets \citep{Tarter07.0}. 
Understanding the evolution of planets on M dwarfs and their potential for hosting life requires 
good knowledge of the stellar properties, and in particular of the phenomena related to
magnetic activity. 
The fact that the habitable zone (HZ) of planets is closer to the star for M stars with respect to 
earlier spectral types implies that planets in the HZ of M stars receive enhanced 
high-energy radiation. 
The stellar emission from the ultraviolet (UV) to the X-ray regime has been recognized to 
(i) critically affect planet evolution and (ii) 
influence the chemistry in the planet atmospheres in a way that is crucial for life.
 
(i) X-ray ($2-100$\,\AA) and EUV ($100-900$\,\AA) radiation lead to heating and 
expansion of a hydrogen-rich planet atmosphere which may become unstable and evaporate 
\cite[e.g.][]{Lammer03.0,CecchiPestellini09.1}. 
Hydrodynamic models for the Earth thermosphere show that strong EUV radiation of the early
Sun may have been able to induce atmospheric blow-off \citep{Tian08.0}. 
The amount by which the planet mass is reduced over time by atmospheric escape clearly 
depends on the radiation input and its time evolution.

(ii) As a consequence of the exposure to strong and variable high-energy radiation 
from the host star the liquid-water HZs defined in \cite{Kasting93.0} 
may not necessarily be `habitable'. 
The most biologically relevant high-energy emission is the near-UV range 
($\sim 2000-3000$\,\AA). 
The appropriate amount of near-UV radiation for a HZ is confined between a minimum flux 
required for biochemical processes to take place and a maximum flux that prevents damage
to biological systems \citep{Buccino06.0}. 
The far-UV ($< 2000$\,\AA) is also important for photochemistry on exoplanets 
\citep{Segura05.0}. 

To date any quantitative evaluation of the evolution of the planets of 
M dwarfs and the possibility of the existence of life on them is hampered by the poorly 
constrained knowledge on the spectral energy distribution of the stellar high-energy 
radiation, the range of activity levels in M dwarfs and their evolution in time.

The coronal X-ray emission of nearby dMe (flare) stars has been intensively studied 
since the beginning of X-ray astronomy \citep{Kahn79.1, Schmitt88.1}.   
The relation between X-ray and bolometric luminosity, $L_{\rm x}/L_{\rm bol}$, 
was observed to saturate at a level
of $\sim 10^{-3}$ for the most active stars with a downward spread of $2-3$\,dex. The strength of 
the X-ray emission has also been known for a while  
to be correlated with the equivalent width of H$\alpha$ emission
\citep{Hawley96.1}, a tracer for magnetic activity in the chromosphere. 
M stars have little photospheric flux at UV and shorter wavelengths.
Their UV emission is almost entirely produced in the chromosphere and transition region. 
%Therefore,
%its relation with the coronal X-ray emission describes the stratification of the upper atmosphere. 
In practice, very little is known on the UV properties of late-type stars in general and of M dwarfs 
in particular, as the UV is one of the least explored wavelength ranges in astronomy because of 
the strong interstellar absorption in this band. 

As a result of our scarse knowledge on the UV emission of M dwarfs, 
for given planetary systems vastly different mass loss histories have 
been predicted depending on the assumptions of the X-ray and EUV flux 
\citep[e.g.][]{Penz08.0,LecavelierdesEtangs07.0}. 
Similarly, the chemistry of exoplanet atmospheres and their suitability for sustaining
life is hard to determine without a good understanding of the incidence of stellar  
X-ray and UV radiation; e.g., \cite{Smith04.0}. 
The few available observations indicate that active M dwarfs outshine the 
Sun at far-UV wavelengths \citep{Segura05.0}. However, these active flare stars 
are not representative for the whole class of M dwarfs and virtually nothing is known
about the UV emission of weakly active M dwarfs.

Until recently, the most extensive data base for stellar UV emission was the IUE archive
\citep{Byrne80.0}. 
As a result of the limitations of the available instrumentation, 
%up to the present day UV spectra are available for very few M dwarfs, and  
it remains unclear what are the relative contributions of the continuum and of the emission lines 
to the quiescent stellar UV radiation of M dwarfs. 
{\em Hubble Space Telescope} (HST) STIS studies of individual flare stars have revealed rich 
emission line spectra but could not detect any UV continuum emission 
\citep{Pagano00.1,Osten06.1,Hawley07.1}.  
The {\em Galaxy Evolution Explorer} (GALEX) has opened a new opportunity for studying
near- and far-UV emission of astrophysical sources \citep{Martin05.1}. 
According to the above-mentioned spectroscopic studies, the GALEX far-UV band 
%(FUV; $1344-1786$\,\AA) 
is dominated by emission lines formed in the transition region such as the 
Si\,{\sc IV}$\lambda\lambda 1394,1403$\,\AA~ and C\,{\sc IV}$\lambda\lambda 1546,1550$\,\AA~ 
doublets and He\,{\sc II}$\lambda 1640$\,\AA.  
The GALEX near-UV band 
%(NUV; $1771-2831$\,\AA) 
is dominated by iron lines. The by far strongest near-UV emission feature, 
Mg\,{\sc II}\,$\lambda\lambda 2796,2803$\,\AA, is at the border of the GALEX near-UV band 
where filter transmission is low such that the major part of this emission feature is lost.

In spite of the fact that the GALEX satellite is dedicated to extragalactic astronomy, 
its large potential for stellar studies has already been demonstrated. 
In a serendipitous observation of the flare star GJ\,3685\,A, a large flare was observed in
both the near-UV and the far-UV bands \citep{Robinson05.1}. 
UV variability on other stars was discussed by \cite{Welsh06.0, Welsh07.0}.
GALEX observations have also been proposed as a tool for identifying M stars in
nearby clusters \citep{Browne09.0} and young associations \citep{Rodriguez11.0, Shkolnik11.0}. 
Finally, the UV activity of solar-type stars has been studied by \cite{Findeisen11.0}
based on GALEX and Ca\,{\sc II}\,H+K measurements. 

In this paper we examine the UV/X-ray connection of the nearest M dwarfs 
by means of a statistical study combining UV measurements from GALEX with X-ray data from 
{\em ROSAT} and {\em XMM-Newton}. 
This dataset is complemented by published spectroscopy yielding H$\alpha$ emission
and $v \sin{i}$. The latter is a proxy for the efficiency of the stellar dynamo that drives
magnetic activity. The sample and the data are introduced in Sect.~\ref{sect:sample}. 
The activity diagnostics derived from these data are described in Sect.~\ref{sect:act}.
The characteristics of the sample's activity are presented in Sect.~\ref{sect:results}.
Sect.~\ref{sect:discussion} comprises a summary of the results and our conclusions.

\section{Sample and catalog}\label{sect:sample}

Our sample is based on the {\it All-Sky Catalog of bright M dwarfs} published
by \cite{Lepine11.0}, henceforth referred to as LG11. 
This catalog comprises 8889 stars selected from the
SUPERBLINK proper motion survey using as criteria high proper motion 
($\mu > 40\,{\rm mas/yr}$), colors typical for M dwarfs ($V-J > 2.7$) and
a magnitude cutoff ($J < 10$). Giant stars have been eliminated 
using cuts in absolute magnitude and reduced proper motion. We refer to
LG11 for details. 

We have selected all $163$ stars from LG11 within $10$\,pc. Within this distance
all stars in the catalog have trigonometric parallaxes and the majority have
an entry in the {\it Catalog of Nearby Stars}, CNS\,3 \citep{Gliese95.1}. 

LG11 provide optical magnitudes from {\em Tycho}, % ($B_{\rm T}$, $V_{\rm T}$),
optical photographic magnitudes from the USNO-B1.0 catalog \citep{Monet03.0},
and near-infrared (NIR) magnitudes from 2\,MASS \citep{Cutri03.1}. They also list 
X-ray counterparts from the {\em ROSAT} All Sky Survey (RASS) 
and far-UV and near-UV counterparts from GALEX. 
The positions used by LG11 are the ICRS coordinates for the 2000.0 epoch and 
have been extrapolated from
the {\em Hipparcos} catalog where available \citep{vanLeeuwen07.0} and from
2\,MASS for the remaining objects. This implies that due to the considerable
proper motion of these nearby stars the cross-correlation with the RASS and
GALEX catalogs is not complete
even more so as relatively small match radii were used by LG11 
($15^{\prime\prime}$ for the RASS and $5^{\prime\prime}$ for GALEX). 
We have matched the coordinates given by LG11 with various catalogs
taking into account the proper motion. As the proper motion correction of the
coordinates depends on the observing date which is not known a priori this is
a two-step procedure the details of which are given below. Briefly summarized,
in the first step, we searched for counterparts in a given
catalog using a large match radius (on the order of $1^\prime$). 
Then we extracted the observing date for the identified counterparts,
we calculated the expected coordinates at that date and repeated the 
cross-correlation with a smaller match radius corresponding to the positional
precision of the instrument. All matches and table handling was performed
with the Virtual Observatory tool TOPCAT \citep{Taylor05.0}.

\subsection{Photometry}\label{subsect:sample_photo}

\subsubsection{X-rays}\label{subsubsect:sample_xrays}

We searched for X-ray counterparts to the M dwarfs using the RASS Bright Source 
Catalog (BSC), the RASS Faint Source Catalog (FSC), the {\it Second ROSAT Source Catalog of 
Pointed Observations} (2RXP), and the {\it XMM-Newton Serendipitous Source Catalog}
released in 2010 (2XMMi-DR3).

The initial match radius for the RASS catalogs 
is $80^{\prime\prime}$ motivated by the maximum
proper motion expected for any of our targets within the approximately ten years
between the RASS and the epoch 2000.0. The coordinates of all sample stars that 
present a RASS counterpart after this step are
extrapolated back to Oct 1, 1990, representative for the observing date of the
RASS which lasted from Aug - Dec 1990.
The subsequent refined cross-match is performed with a radius of 
$40^{\prime\prime}$ \citep{Neuhaeuser95.1}. 
With this procedure all $58$ X-ray identifications provided by LG11 for the 
163 sample stars were recovered. Moreover, an additional $24$ X-ray 
counterparts not listed as X-ray sources in LG11 were identified 
this way in the RASS. 

For the cross-correlation of the pointed X-ray catalogs, 2RXP and 2XMMi-DR3, 
we performed the proper motion correction of the coordinates individually
for each object and for the given observation date. For the rest, 
we proceeded in the same way as for the RASS.  
%(actually $100^{\prime\prime}$ match radius in the first step!). 
The final match radius for the 2RXP is again $40^{\prime\prime}$ as found to be 
appropriate for pointed PSPC observations by \cite{Stelzer01.1}. 
This way we identified $24$ of the $163$ stars as pointed {\em ROSAT} sources.
Sixteen of them were not listed as X-ray emitters by LG11 
and $9$ of them are not detected in the RASS.
The majority are located within $10^{\prime\prime}$ of the optical/NIR position and
the maximum displacement is $25^{\prime\prime}$. 

In an analogous way, $11$ stars are identified in the 2XMMi-DR3. Although we permitted
a rather large match radius of $20^{\prime\prime}$ after the proper motion 
correction all {\em XMM-Newton} counterparts are within $3^{\prime\prime}$ of the
expected optical coordinates. For one star with two {\em XMM-Newton} identifications
at similar separation from the optical source ($\sim 2^{\prime\prime}$) we 
retained the one from the observation with longer exposure time. Only two stars
with {\em XMM-Newton} detection are not detected in any of the {\em ROSAT} catalogs.
These two are actually one {\em XMM-Newton} source identified with two optical objects,
PM I23318+1956 E + Wn (Gl\,896\,A+B). 

To summarize, $93$ of $163$ M dwarfs within 10\,pc have an X-ray counterpart,
i.e. there are $\approx 57$\,\% more identifications than presented by LG11.

\subsubsection{Ultraviolet}\label{subsubsect:sample_uv}

{\it GALEX} performs imaging in two UV bands,
far-UV (henceforth FUV; $\lambda_{\rm eff} = 1539$\,\AA, $\Delta \lambda = 1344-1786$\,\AA)
and near-UV (henceforth NUV; $\lambda_{\rm eff} =2316$\,\AA, $\Delta \lambda = 1771-2831$\,\AA)
with a spatial resolution of $4.2^{\prime\prime}$ and $5.3^{\prime\prime}$
in the two bands, respectively.
Three nested {\it GALEX} imaging surveys
have been performed: the All-Sky Survey (AIS)  covering
a large fraction ($\sim 85$\,\%) of the high Galactic latitude ($\|b\|> 20^\circ$)
sky to $m_{\rm AB} \sim 21$\,mag, the Medium Imaging Survey (MIS) reaching
$m_{\rm AB} \sim 23$\,mag on 1000 deg$^2$, and the Deep Imaging Survey (DIS)
extending to $m_{\rm AB} \sim 25$\,mag on 80 deg$^2$ \citep[e.g.][]{Bianchi09.0}. %, Bianchi11.0}.
These main surveys are complemented by guest investigator programs.

The $10$-pc sample was cross-correlated with the {\it GALEX} data distributed in the
Data Release 6 (GR6), which has been homogeneously reduced
and analysed by a dedicated software pipeline, available at the Mikulski Archive for Space Telescopes
(MAST)\footnote{http://galex.stsci.edu/casjobs}.

In the first step of our cross-correlation procedure,
we used the maximum possible match radius in MAST ($60^{\prime\prime}$).
After the proper motion correction for the GALEX observing dates extracted for
our targets, we reduced the match radius to
$10^{\prime\prime}$ for the final cross-correlation.
Only sources having a distance
from the center of the field of view $\leq 50^\prime$ were retained
as generally the photometric quality is better in the
central part of the field \citep{Bianchi11.0}.
In the case of multiple matches in the same observation, 
we retained the nearest source to the target. If the multiple matches were
from different observations, they were assumed to be multiple
observations of the same source.
This yields {\it GALEX} counterparts to $72$ of the $163$ sample stars,
of which all but one were observed in both FUV and NUV.
All $71$ stars observed in the NUV were detected while $49$ of the $72$ stars observed
in the FUV were detected. 
Most of the stars (82\%) were observed in the AIS, $11$ were observed in the MIS  
and only $2$ in the DIS. 
Fifty six ($\sim$ 34\%) of the remaining $91$ stars of the sample,
do not fall in the area covered by {\it GALEX}.

We note that only $30$ stars are listed as NUV and $54$ listed
as FUV sources in LG11. Analogous to the case of the X-ray sources, our significantly
larger number of UV identifications is likely 
due to the fact that LG11 have not corrected for proper motion when cross-correlating
catalogs.

\subsection{Spectroscopic parameters}\label{subsect:sample_spec}

LG11 provide spectral types for all sample stars derived from 
the $V-J$ color using a calibration based on SUPERBLINK stars that have a 
{\em Sloan Digital Sky Survey} (SDSS) 
spectrum. In their follow-up work of their proper-motion selected catalog,
\cite{Lepine12.0} 
%Lepine et al. (2012, AJ submitted)  
checked that calibration for a spectroscopically confirmed part of the sample. 
They found that for stars
with red colors this $V-J$ calibration tends to predict spectral types that are 
between $1-2$ subclasses too cool. They redetermined the $V-J$ vs. spectral
type relationship
by fitting a third-order polynomial to the observed distribution. 

We adopted this updated calibration to compute the spectral types for all
stars in the $10$-pc sample from the $V-J$ colors given by LG11. 
We performed an independent literature search and found spectroscopically derived 
spectral types for all but two stars of the $10$-pc sample. 
In particular, we have cross-matched our sample with
the recent catalog of rotation and activity in early-M stars by \cite{Reiners12.0},  
with \cite{Mohanty03.1} which comprises stars with spectral type M4 and later
and with the Palomar/MSU survey by \cite{Reid95.2}. These three catalogs together
yield spectral types for $154$ stars. Another $7$ have spectral types listed in
\cite{RojasAyala12.0, Torres06.0, Scholz99.0, Reid07.1}. 

\begin{figure}
\begin{center}
\includegraphics[width=9cm]{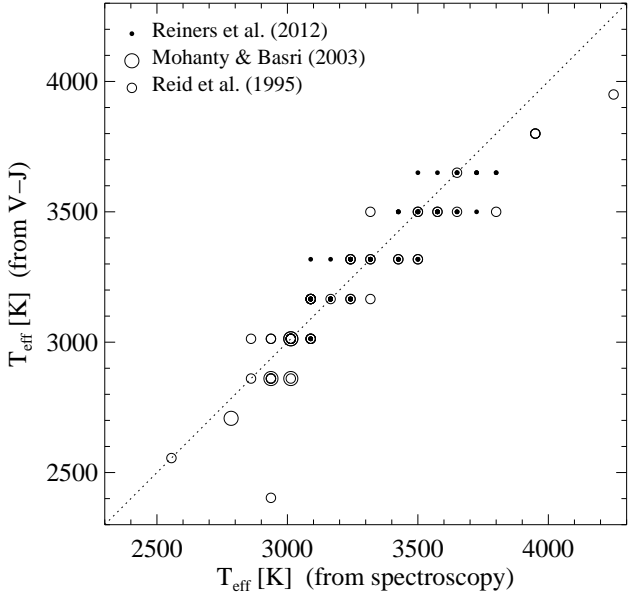}
\caption{Comparison of the effective temperatures obtained from the $V-J$
vs. spectral type calibration given by \protect\cite{Lepine12.0} and the spectral types from the 
spectroscopic literature. The adopted temperature scale is from   
\protect\cite{Bessell91.1} and \protect\cite{Mohanty03.1}.  
Several of the $159$ stars in this plot are at the same position. 
Omitted are only the two stars without spectroscopic spectral
type determination and the two stars with spectral type cooler than M9 
according to \protect\cite{Lepine12.0}.} 
\label{fig:teff_comparison}
\end{center}
\end{figure}

To obtain the effective temperature ($T_{\rm eff}$) from the spectral types
we combined the temperature scales from \cite{Bessell91.1} and \cite{Mohanty03.1}.
These two scales span the full M spectral class. 
A comparison between the effective temperatures obtained from the spectroscopically
determined spectral types with those from $V-J$ is shown in Fig.~\ref{fig:teff_comparison}.
The corrected $V-J$ vs. spectral type 
relation presented by \cite{Lepine12.0} 
is in good agreement with the 
spectroscopically determined spectral types with few exceptions. These comprise
also two stars for which the $V-J$ vs. spectral type relation predicts 
spectral types cooler than M9 and that are not covered by our temperature scale. 
The spectroscopic data provides spectral types with accuracy down to $\pm 0.5$
subclass while the spectral types derived from the photometry have been rounded
to $\pm 1$ subclass. 

We adopted throughout this paper the spectral types from the spectroscopic literature,
except for the two stars without available spectral types for which we resort to 
the values derived from $V-J$ as described above. 
Moreover, we have removed four stars with final spectral classification late K
such that the $10$-pc sample is reduced to $159$ stars. 

%
% OUTPUT FROM  compare_teff.pro
%

\begin{figure*}
\begin{center}
\parbox{18cm}{
\parbox{6cm}{
\includegraphics[width=6cm,angle=0]{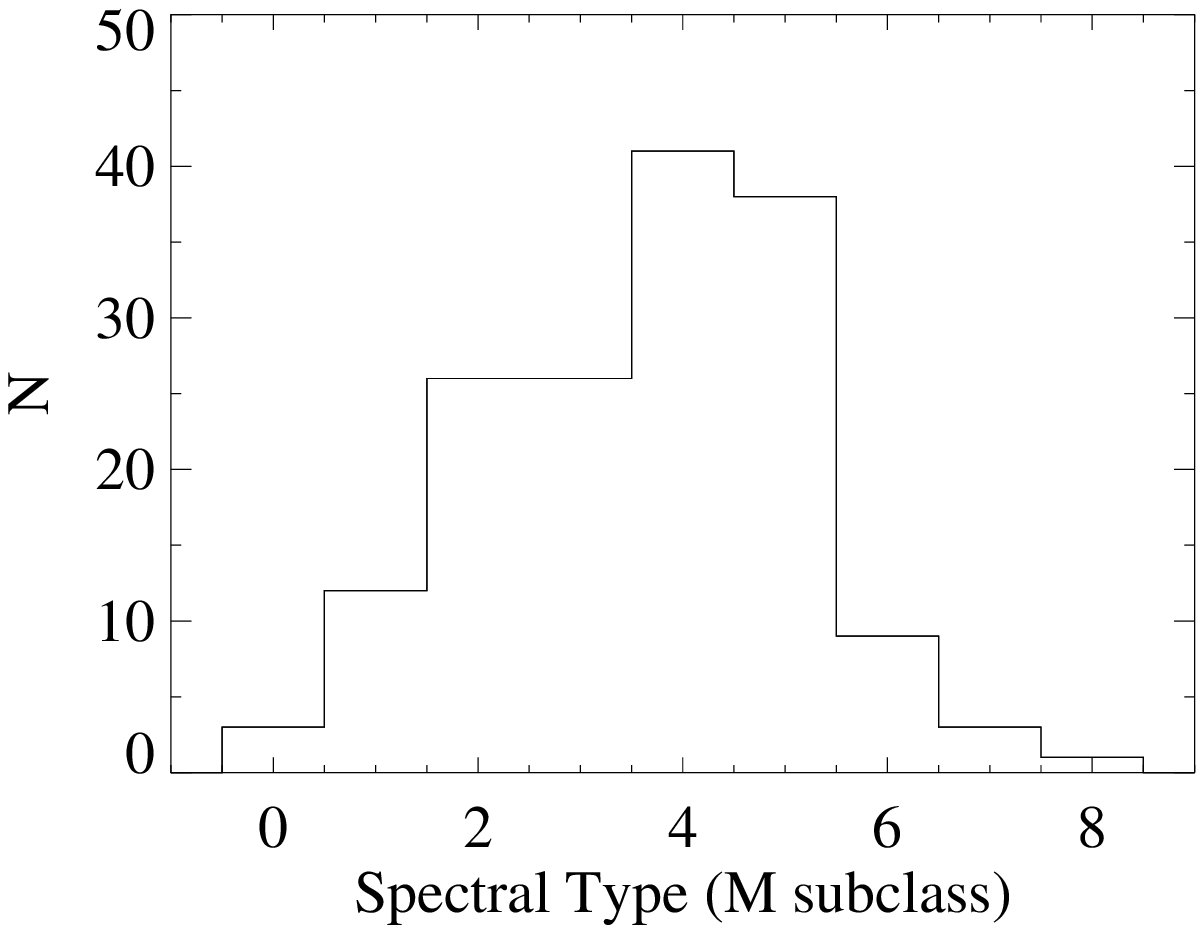}
}
\parbox{6cm}{
\includegraphics[width=6cm,angle=0]{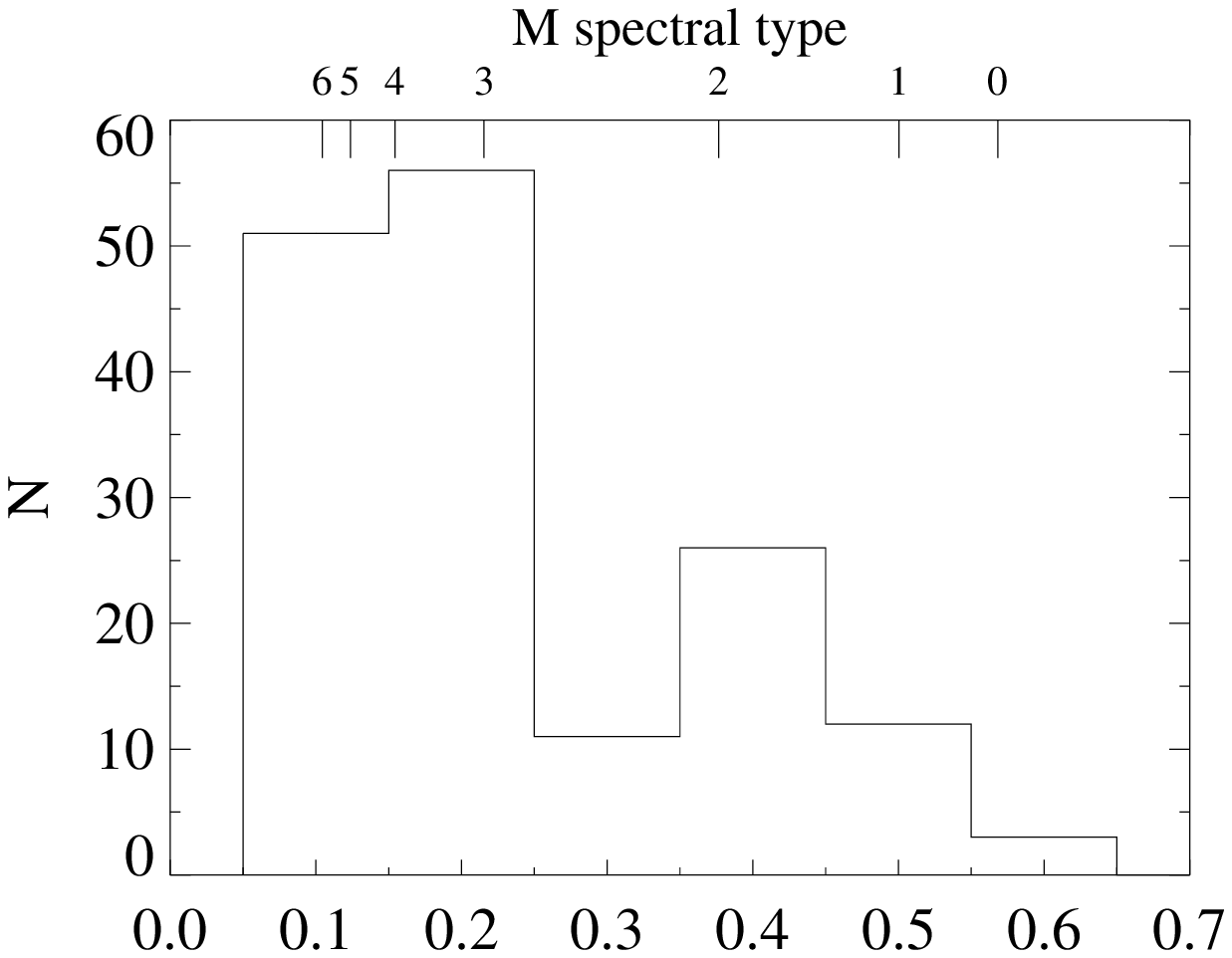}
}
\parbox{6cm}{

\includegraphics[width=6cm,angle=0]{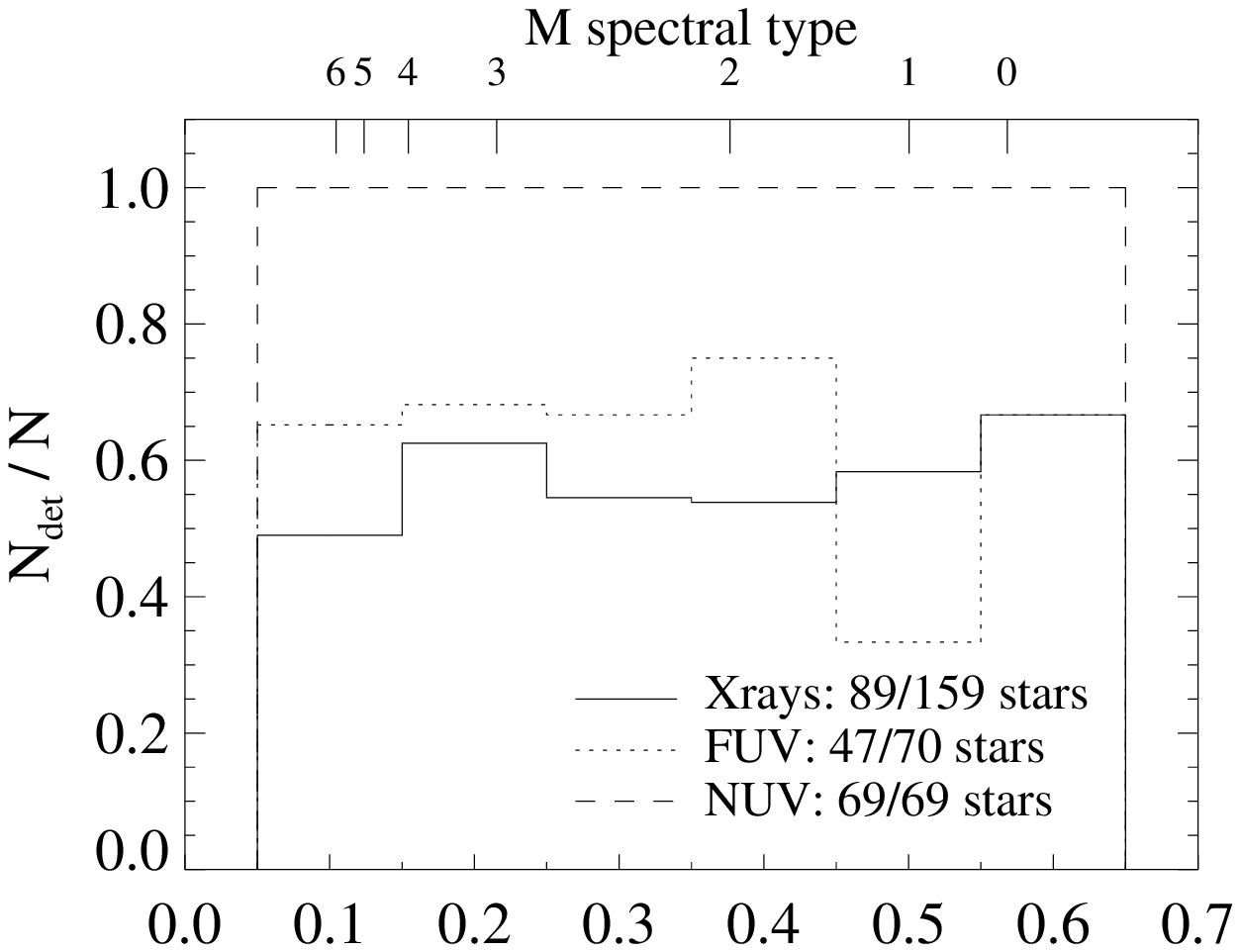}
}
}
\caption{Spectral type (left panel) and mass (middle panel) distribution of the 
$10$\,pc-sample. The fractions of the subsample in each mass bin that are detected in 
X-rays (solid line), FUV (dotted line) and NUV (dashed line) 
are shown in the right panel.}
\label{fig:histo_spt}
\end{center}
\end{figure*}

Fig.~\ref{fig:histo_spt}\,a and~b display the spectral type and the mass 
distribution of the final $10$\,pc-sample. 
The masses of all stars were derived from their $T_{\rm eff}$ making use of 
the $5$\,Gyr isochrone of the \cite{Baraffe98.1} models. 
The two histograms in Figs.~\ref{fig:histo_spt}\,a and~b are very different 
reflecting the non-linear
transformation between temperature and mass. 
In the right panel of the figure the mass distribution of the subsamples detected
in X-rays, FUV and NUV, respectively, is displayed. 
Note that the number of stars with observations and that with detections in one
of the three energy bands differs from the numbers given in 
Sects.~\ref{subsubsect:sample_xrays}
and~\ref{subsubsect:sample_uv} due to the removal of four K-stars from the sample. 
In each mass bin the number
of detected stars in a given energy band has been normalized to the total number 
of stars in that mass bin with available data in the same energy band. A detection
fraction of $100$\,\% is reached only in the NUV. 
For the other two bands, X-rays and FUV, 
the detection fraction ranges between $\sim 30-80$\,\% and 
is remarkably constant as a function of mass. 
This means that with the presently available instrumentation it is not
possible to reach down to the X-ray and FUV emission levels of the least active 
nearby M stars. 
%
% OUTPUT FROM   plot_histo_spt_mass.pro
%

The spectroscopic catalogs of \cite{Reiners12.0} and \cite{Mohanty03.1} 
provide, next to the spectral types, information on the rotational velocity
($v \sin{i}$) and H$\alpha$ emission. 
Within the $10$-pc sample, $67$ stars have an H$\alpha$ measurement, 
$25$ of them are detections and $42$ are upper limits. 
A measurement for the rotational velocity has been done for $107$ stars from 
the spectroscopic catalogs but for $83$ of them this yielded 
only an upper limit to $v \sin{i}$. 

The relevant stellar parameters of all $159$ stars from 
the $10$-pc sample are given in Table~\ref{tab:tenpc_params_ste}. Next to the identifiers 
from LG11 (col.1) we give in col.2 the Gl/GJ numbers from the CNS.  
The subsequent columns represent
the distances corresponding to the trigonometric parallaxes given by LG11, 
the spectral types and effective temperatures as described above and the bolometric 
flux at Earth, $f_{\rm bol}$. 
This latter one is obtained from 
the surface flux, $F_{\rm s,bol} = \sigma_{\rm B} T_{\rm eff}^4$
where $\sigma_{\rm B}$ is the Boltzmann constant, multiplying it with the
scaling factor $(\frac{R_*}{d})^2$. 
The stellar radii ($R_*$) were derived from \cite{Baraffe98.1}
models assuming an age of $5$\,Gyrs for all stars. 
The $\pm 0.5$ subclass 
uncertainties in spectral type correspond to about $\pm 75$\,K in temperature.
According to \cite{Baraffe98.1} the stellar radius changes by less than $3$\,\% 
between $1-5$\,Gyrs, the likely age of this sample. 
The errors in the distances are for most stars also very small. Therefore, the 
uncertainties in $f_{\rm bol}$ are dominated by the uncertainty in the temperature
and amount to circa $\approx 10$\,\%.

\subsection{Completeness}\label{subsect:sample_complete}

To examine the completeness of the $10$-pc sample we followed two approaches.
First, we compared the sample to the previous samples of nearby M stars presented in 
the literature. 
Secondly, we computed the mass function of the $10$-pc sample and compared it
to the mass function of field M dwarfs presented by \cite{Bochanski10.0}. 

We noticed that $10$ out of $39$ stars of 
the list of M dwarfs within $10$\,pc studied by \cite{Marino00.1} are missing in the
$10$-pc sample. This includes some of the most well-known flare stars, 
UV\,Cet, CN\,Leo, and Prox\,Cen. The absence of one star from
the $10$-pc sample can be attributed to it being at the border of our distance
cutoff (with $d = 11.4$\,pc in LG11 but $d <10$\,pc in \cite{Marino00.1}). 
The remaining $9$ missing stars 
imply a completeness of $95$\,\% for the $10$-pc sample. LG11 consider their
full (not distance-limited) M dwarf catalog $\sim 90$\,\% complete for the
northern hemisphere and $\sim 60$\,\% for the southern hemisphere. This is
qualitatively consistent with the fact that 
only one of the $9$ stars missing from the $10$-pc sample has positive declination. 
Assuming that there are no other stars missing in the $10$-pc census of LG11 
beyond these $9$ objects, we find a completeness of $99$\,\% for the northern 
and of $89$\,\% for the southern sky.

\begin{figure}
\begin{center}
\includegraphics[width=9cm,angle=0]{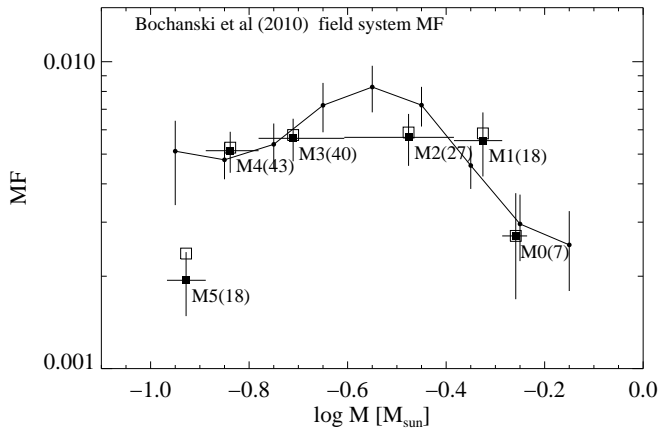}
\caption{Space density of the $10$-pc sample (filled squares) compared to the M dwarf 
mass function from \protect\cite{Bochanski10.0} (shown as small circles connected
by a solid line). Numbers in brackets indicate the number of
stars in each spectral type bin for the $10$-pc sample. 
Open squares represent the $10$-pc sample incremented
by the nine missing stars from \protect\cite{Marino00.1}.}
\label{fig:mf}
\end{center}
\end{figure}

\cite{Bochanski10.0} have presented the mass function of low-mass field dwarfs 
based on SDSS photometry for $15$\,million stars. 
In Fig.~\ref{fig:mf} we reproduce their result for the system mass function (small
circles connected by the solid curve). 
Over-plotted is the space density of the $10$-pc sample divided into spectral subclasses. 
Each spectral type bin combines stars with spectral type MX and MX.5, 
where ${\rm X} = 0...5$, and is centered on MX.25. The spectral types were 
converted to mass with the temperature scale
and evolutionary models as described in Sect.~\ref{subsect:sample_spec}.
We omitted spectral bins at the edge of the samples' mass distribution that are populated
by less than $5$ stars. In practice this regards 6 stars with spectral types M6 and later as 
the outliers of late-K type have already been removed from the sample.   
The $10$-pc sample is overall in good agreement with the mass function presented by
\cite{Bochanski10.0} except for the coolest mass bin where a large number of
stars seem to be missing. 
Assuming that we miss half of the stars in the M5 bin, the number necessary to
make our sample comply with the mass function of \cite{Bochanski10.0} at M5 spectral type, 
our completeness is $153 / (153 + 18)$, i.e. $\sim 89$\,\%. 
The `missing stars' from \cite{Marino00.1}, indeed, have a peak in this bin
(see open squares in Fig.~\ref{fig:mf}). 
%
%  OUTPUT FROM   massfunction.pro
%

\section{Activity diagnostics}\label{sect:act}

\subsection{X-ray fluxes}\label{subsect:act_xrays}

First, the X-ray data compiled from the various catalogs
that we have cross-correlated with the $10$-pc sample must be brought
into a homogeneous shape. 
The RASS catalogs and the 2RXP list count rates in the {\em ROSAT}/PSPC energy band 
and the 2XMMi-DR3 catalog gives fluxes in various energy bands between $0.2-12$\,keV. 
These fluxes refer to a power-law spectrum with index $\Gamma = 1.7$ 
absorbed by $N_{\rm H} = 3 \cdot 10^{20}\,{\rm cm^{-2}}$ \citep{Watson09.1}. 
We have extracted the $0.2-2.0$\,keV fluxes assuming a thermal one-temperature
spectrum from all X-ray data described in Sect.~\ref{subsubsect:sample_xrays} as follows. 

Concerning the 2XMMi-DR3 data, 
we must note that a power-law is representative for X-ray emission generated by 
non-thermal electrons and is not adequate for describing the thermal coronal X-ray 
emission. 
Moreover, the interstellar absorption of nearby stars is much smaller than 
assumed in the 2XMMi-DR3. We calculated the count-to-flux conversion factors ($CFs$)
for a thermal one-temperature model (APEC) of $kT = 0.3$\,keV\footnote{Our estimate of the coronal temperature is based on previous X-ray studies of
late-type stars, e.g. for solar analogs there is an empirical power law relation
between X-ray luminosity and temperature with $T \sim 2-6$\,MK for 
$\log{L_{\rm x}}\,{\rm [erg/s]} = 27...29$ \citep{Guedel04.0}. This corresponds  
approximately to the range of X-ray luminosities of the $10$-pc sample.} and a column density
of $10^{19}\,{\rm cm^{-2}}$, corresponding to the average interstellar
extinction at $10$\,pc. 
The fluxes in the 2XMMi-DR3 energy bands were then
corrected using the ratio between the (power-law) $CFs$ used in the {\em XMM-Newton}
Serendipitous Source Catalog\footnote{The count-to-flux conversion factors used in the 
XMM-Newton Serendipitous Source Catalog are available at http://xmmssc-www.star.le.ac.uk/Catalogue/2XMM/UserGuide\_xmmcat.html\#Tab ECFs} 
and the APEC $CFs$ calculated by us. Finally,
we summed up the corrected 2XMMi-DR3 fluxes from the three lowest energy bands 
to obtain the $0.2-2.0$\,keV flux. 
(For better comparison with {\em ROSAT} fluxes we do not consider the higher
{\em XMM-Newton} energy bands from $2.0-4.5$\,keV and from $4.5-12$\,keV.) 
The PSPC count rates from the {\em ROSAT} catalogs are translated into
$0.2-2.0$\,keV fluxes 
using the same thermal model as 
for the {\em XMM-Newton} data 
($CF_{\rm ROSAT} = 2.03 \cdot 10^{11}\,{\rm cts/erg/cm^2}$). 
%($CF_{\rm ROSAT} = 1.55 \cdot 10^{11}\,{\rm cts/erg/cm^2}$). 
%CF=2.03d11  for kT=0.3, N_H=1e19
%CF=8.19d10  for kT=1, N_H=3e20
%CF=1.545d11  for kT=1, N_H=1e19

\begin{figure}
\begin{center}
\includegraphics[width=9cm,angle=0]{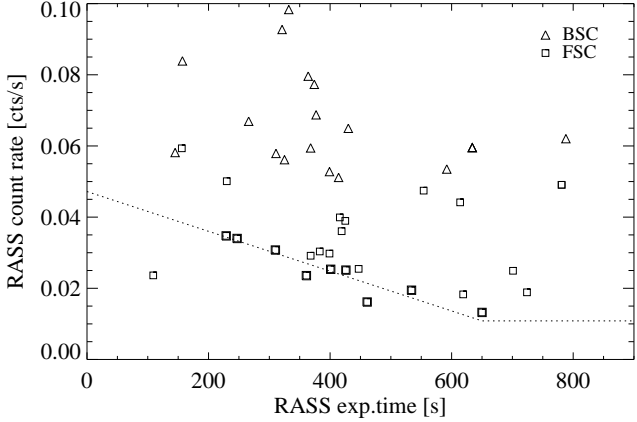}
\caption{Count rate for stars from the $10$-pc sample from the RASS
BSC (triangles) and FSC (squares) as function of the exposure time.
The dotted line is the empirical lower envelope that defines
the sensitivity limit. The linear part was determined by fitting the stars
represented by thicker plotting symbols. For exposure times larger than 
$650$\,s we assumed a constant upper limit count rate.}
\label{fig:rass_sensitivity}
\end{center}
\end{figure}

About $40$\,\% of the stars in the $10$-pc sample have no X-ray detection,  
and we calculated their upper limits after having extracted the sensitivity
limit making use of the RASS detections as follows. 
Fig.~\ref{fig:rass_sensitivity} shows those stars from the $10$-pc sample 
that were detected in the RASS BSC or FSC. The lower envelope of the distribution 
of count rate versus exposure time defines the RASS sensitivity threshold. We
obtained a numerical value for the count rate upper limit as a function of the
RASS exposure time by performing a linear fit to the lowest observed count rates for
given exposure time (dotted line in Fig.~\ref{fig:rass_sensitivity}).
For large exposure times a constant sensitivity limit is assumed. 
This is a conservative assumption considering that for long observations the
sensitivity becomes background limited and decreases as the square root of the
exposure time. 
We extracted the RASS exposure time for the proper motion corrected ICRS position of each of
the undetected stars from the 
HEASARC archive\footnote{http://heasarc.gsfc.nasa.gov/cgi-bin/W3Browse/w3browse.pl}, 
and placed them on the sensitivity curve to obtain their upper limit count rate. 
Then $CF_{\rm ROSAT}$ was applied for the conversion to the $0.2-2.0$\,keV flux. 

\begin{figure}
\begin{center}
\includegraphics[width=9cm,angle=0]{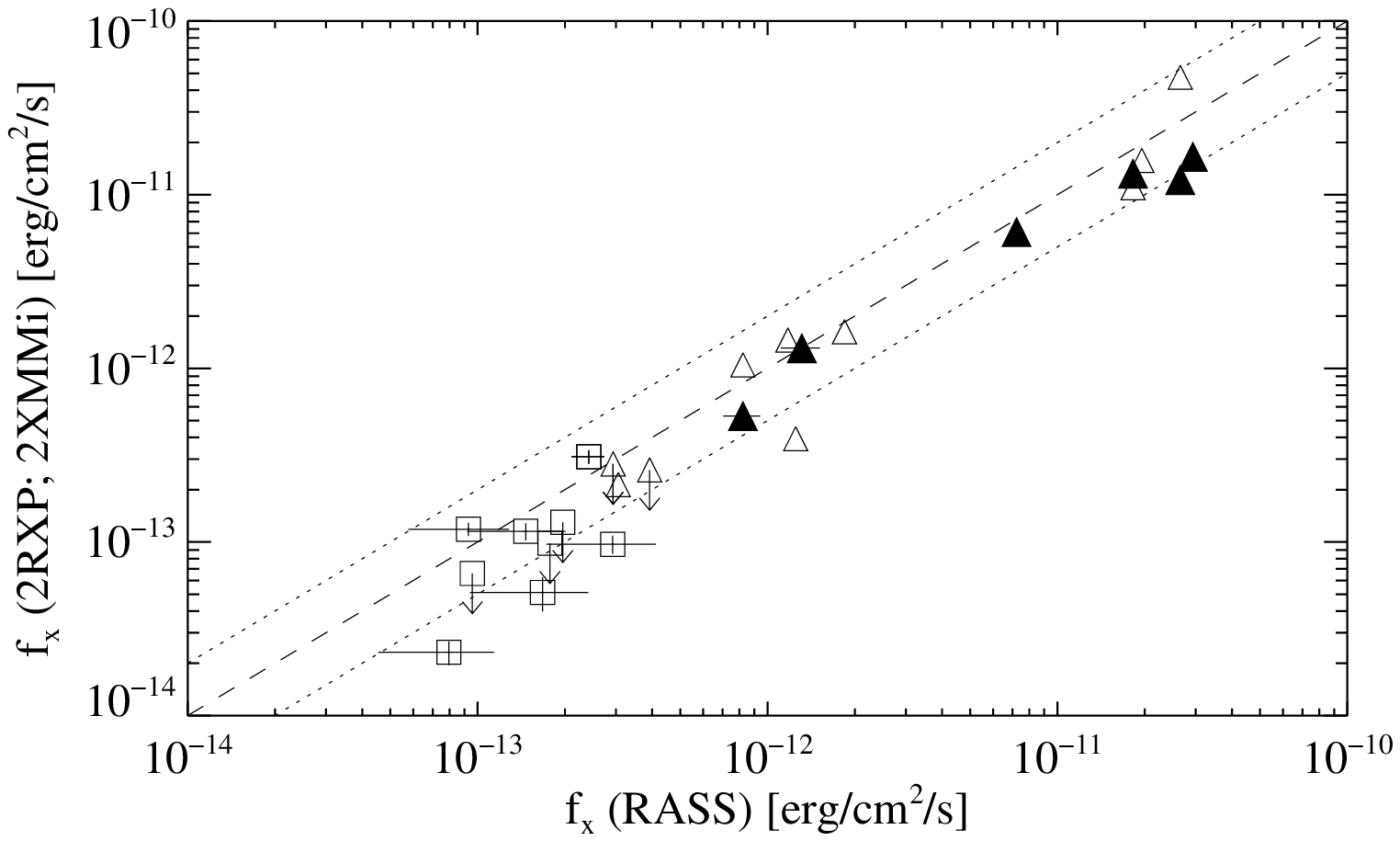}
\includegraphics[width=9cm,angle=0]{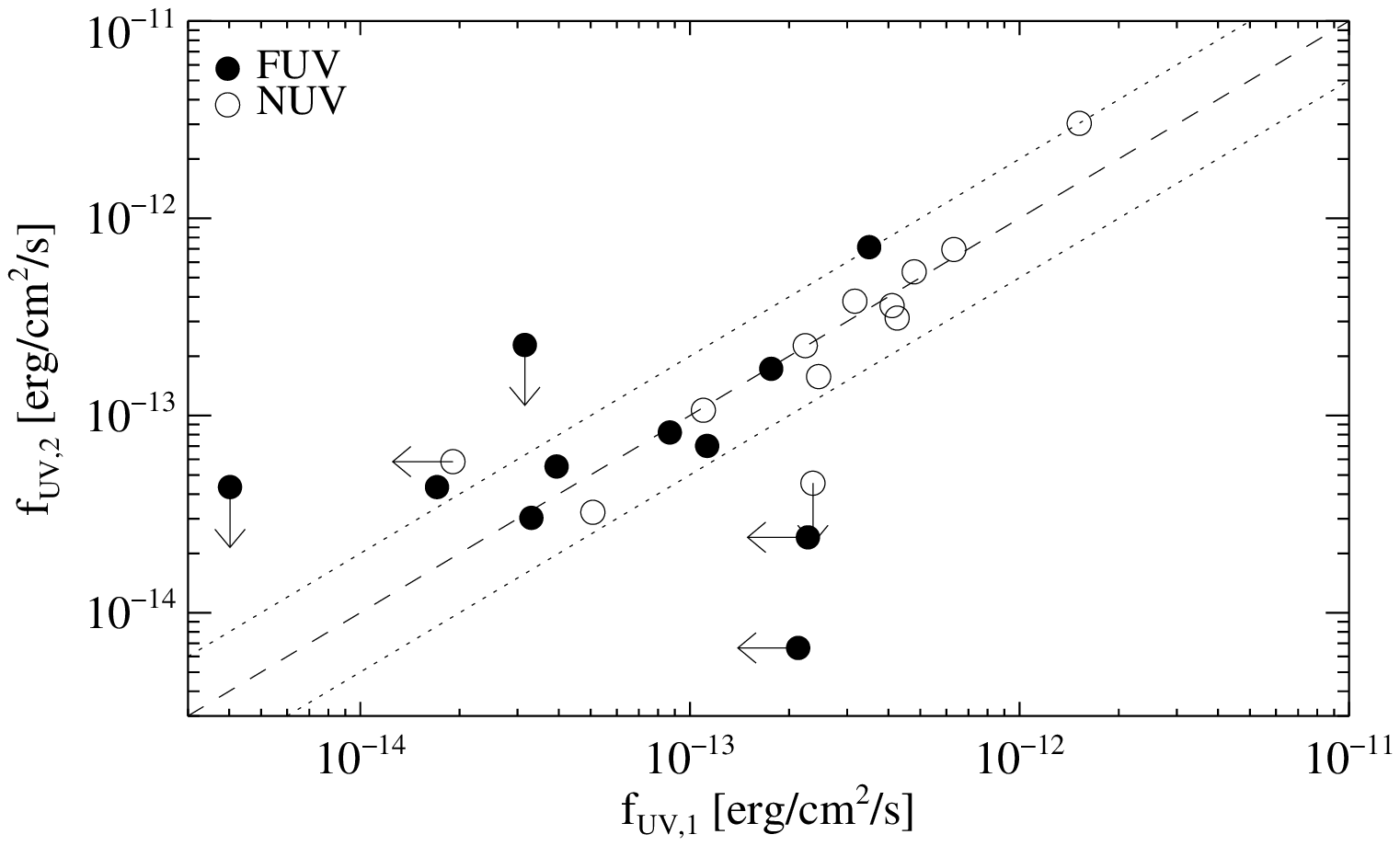}
\caption{Comparison of X-ray and UV flux from multiple observations of a given star.
(top) -- pointed X-ray data versus RASS for the stars with multiple X-ray 
detections; (bottom) -- both FUV (filled circles) and NUV (open circles).}
\label{fig:variability}
\end{center}
\end{figure}

% OUTPUT FROM check_xraydata.pro  for sample with 159 stars (4 K stars removed):
% 27 pairs of X-ray observations for 24 distinct stars; 3 have 3 observations;
% figure pointed_vs_rass shows 25 data points because 2 X-ray sources are identified
% with 2 stars each: 06337-7537N+S, 18427+5937N+S
There are $25$ stars in the $10$-pc sample that are reported in more than one
of the X-ray catalogs that we have searched, allowing us to look for evidence
of variability. For those stars, the X-ray flux from the pointed observations is 
compared to the X-ray flux from the RASS in Fig.~\ref{fig:variability}.
The majority of stars do not display variations larger than about a factor of
two. This is consistent with an earlier variability study of M dwarfs by \cite{Marino00.1}. 
Having verified the negligible role of variability, we can consider a single measurement 
for each star to be representative for its X-ray emission level. We selected
the flux from the 2RXP if available %($25$ stars), 
else the flux from the BSC %($48$ stars) 
or the FSC. %($17$ stars). 
Only if no {\em ROSAT} detection is available we resort to the 2XMMi. % ($2$ stars). 
For the %$64$ 
undetected stars we use the RASS upper limits. 

The X-ray fluxes and upper limits of all stars from the $10$-pc sample are listed in 
Table~\ref{tab:tenpc_params_act} together with their uncertainties in squared brackets.  
For further analysis, the X-ray fluxes and their upper limits were transformed into 
luminosities and their upper limits 
using the individual distances listed in Table~\ref{tab:tenpc_params_ste}. 

\subsection{UV fluxes}\label{subsect:act_UV}

FUV and NUV magnitudes in the AB-system 
within a 12.8$^{\prime\prime}$ diameter of the source position 
were extracted for all detections in the $10$-pc
sample. This diameter is a good compromise between the large apertures required
to gather almost all of the {\it GALEX} flux  and the small ones that are optimal for minimising
the background. 
Flux densities were computed from the FUV and NUV magnitudes using the
transformation 
relations given by \cite{Bianchi09.0}, and they were converted to fluxes with the
effective band width\footnote{see GALEX Observer's guide available at http://galexgi.gsfc.nasa.gov 
/docs/galex/Documents/ERO\_data\_description\_2.htm} 
of the respective GALEX filter ($\delta \lambda_{\rm FUV} = 268$\,\AA; 
$\delta \lambda_{\rm NUV} = 732$\,\AA). 
%\cite{Morrissey07.0}. 

The upper limits of the 
% ANTO: $26$ 
% my numbers in reduced 10pc sample: $23$
stars that are undetected in the FUV band were estimated as follows.
We obtained the FUV upper limit of PM\,I03133+0446S (GJ\,1057), the only star undetected in the MIS,
using the background-subtracted FUV {\it GALEX} intensity image. We computed the FUV magnitude
of PM\,I03133+0446S within a circular aperture centered on the star and with the same diameter 
as the aperture used for the detected sources. 
All remaining undetected stars were observed in the AIS. We estimated 
the FUV upper limits for $5$ of them, spanning all the range of exposure times, 
in the same manner as for PM\,I03133+0446S. The upper limits of the others 
were then obtained by linear interpolation between these values.

There are $12$ stars with two NUV observations and $11$ stars with two FUV observations. 
These objects can be investigated for UV variability analogous to the case of the X-ray emission
in Sect.~\ref{subsect:act_xrays}. The bottom panel of 
Fig.~\ref{fig:variability} compares the fluxes for stars with two measurements 
in the FUV and/or the NUV band. Overall  
there is little evidence for variability. Only the upper limits are far off from 
the 1:1 relation. However, all FUV upper limits are compatible with the 
detections, and they are shallow due to the low exposure time in the respective 
observation. The only star that has shown strong variability is PM I09307+0019
(GJ\,1125). 
Its NUV upper limit is more sensitive than its NUV detection. 
This is not a contradiction because the exposure times of its two observations
were different by about a factor 10. 
In the remainder of this paper, we used for stars that are detected in two
observations of a given GALEX band, the average of the two measurements.
Observed but undetected stars are represented by their upper limits. 
We ignore the upper limits of stars that also have a detection in another
GALEX observation. 

Analogous to the X-ray fluxes, we list the FUV and NUV fluxes in 
Table~\ref{tab:tenpc_params_act}. 
Luminosities in the two GALEX bands were calculated from the fluxes in the same way as
for the X-ray band (see Sect.~\ref{subsect:act_xrays}).

\subsection{H$\alpha$ fluxes}\label{subsect:act_ha}

The spectroscopic catalogs described in Sect.~\ref{subsect:sample_spec} give the H$\alpha$
to bolometric flux ratio ($\log{(f_{\rm H\alpha}/f_{\rm bol})}$) that we designate `activity
index' in Sect.~\ref{subsect:act_rindex}. We have extracted the observed fluxes from this 
quantity reversing the approach for obtaining the UV activity index described next.

\subsection{Activity indices}\label{subsect:act_rindex}

In analogy to the standard measure for Ca\,{\sc II}\,H+K activity, we define the 
UV activity index as 
\begin{equation}
R^\prime_{\rm UV} = \frac{f_{\rm UV,exc}}{f_{\rm bol}} = \frac{f_{\rm UV,obs} - f_{\rm UV,ph}}{f_{\rm bol}} 
\label{eq:act_index}
\end{equation}
where $f_{\rm UV,exc}$ is the UV excess flux attributed to activity, 
i.e. the difference between the observed UV flux ($f_{\rm UV,obs}$) and the 
photospheric flux in the same UV band ($f_{\rm UV,ph}$), % predicted by the bestfit model, 
and $f_{\rm bol}$ is the bolometric flux. % derived from the SED fitting. 
The bolometric flux is obtained from 
the surface flux as described in Sect.~\ref{subsect:sample_spec}. 

\begin{figure*}
\begin{center}
\parbox{16cm}{
\parbox{8cm}{
\includegraphics[width=8cm,angle=0]{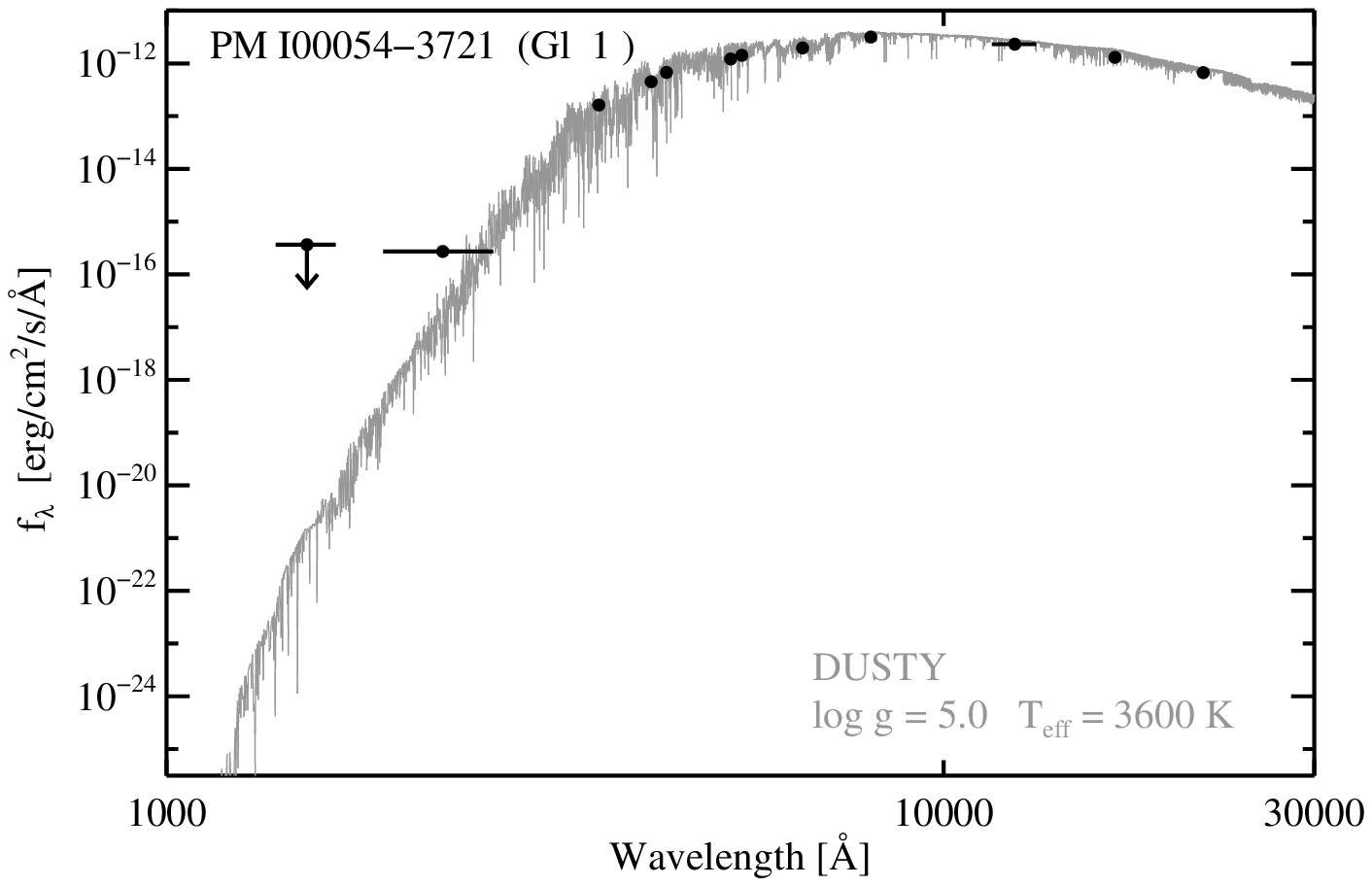}
}
\parbox{8cm}{
\includegraphics[width=8cm,angle=0]{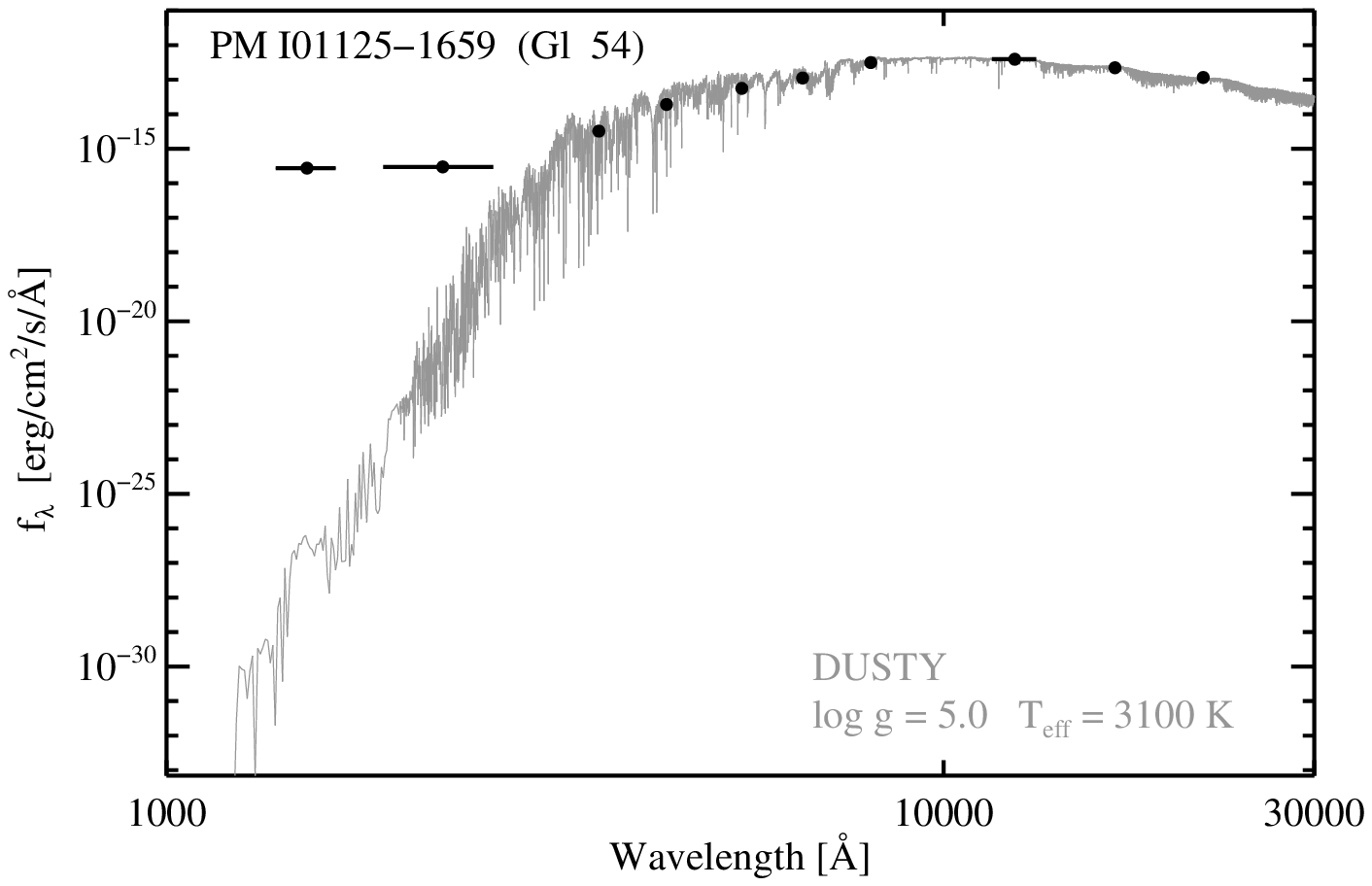}
}
}
\caption{Examples of observed SEDs for two stars from the $10$-pc sample compared to
the DUSTY model spectrum selected as described in Sect.~\ref{subsect:act_rindex}. The
photometry was compiled from data archives and catalogs: 
{\em Tycho-2} and {\em Hipparcos} \protect\citep{Hog00.1, Perryman97.1},
SDSS\,DR7 \protect\citep{Adelman09.0}, 
2\,MASS \protect\citep{Cutri03.1}
and multi-band photometry from \protect\cite{Koen10.0}. 
The two left-most data points represent the GALEX FUV and NUV bands.}
\label{fig:sed_examples}
\end{center}
\end{figure*}

In Eq.~\ref{eq:act_index} `UV' stands for the NUV and FUV bands, respectively.
The superscript ($^\prime$) indicates that the ratio has been corrected for
photospheric emission. 
The photospheric contribution to the NUV and FUV emission was estimated with help
of the synthetic DUSTY spectra of \cite{Allard01.1}. 
%The observed NUV and FUV fluxes which include photospheric and chromospheric contributions   
%are to be compared to the surface fluxes of the DUSTY spectra representative of the photosphere. 
We adopted for each star in the
$10$\,pc sample the model spectrum with solar metallicity, $\log{g} = 5$ and with 
$T_{\rm eff}$ corresponding to the spectral type from the references described in 
Sect.~\ref{subsect:sample_spec}. 
Fig.~\ref{fig:sed_examples} shows two examples of observed spectral energy distributions
(SEDs) and corresponding DUSTY model. 
The synthetic spectra are available in terms of surface flux density 
and have been transformed for this plot   
into observed flux density by applying the dilution factor $(\frac{R_*}{d})^2$.
As before in Sect.~\ref{subsect:sample_spec} 
we have used for the stellar radius the value extracted from the $5$\,Gyr isochrone 
of \cite{Baraffe98.1} for the known effective temperature of each star.
As mentioned in Sect.~\ref{subsect:sample_spec}, no significant difference in radius is
predicted for ages between $1$ and $5$\,Gyr. 
% 
% OUTPUT FROM   plot_SED_from_modelspectra.pro
%

For many stars, when comparing the observed flux densities to the model spectra as
described above we noticed serious problems with the normalization. If the 
distance is assumed to be known from the trigonometric parallax for each star, 
for those cases an unrealistically large radius was needed to make the model 
spectrum fit the observed photometry. A larger radius corresponds for given $T_{\rm eff}$ to 
a smaller gravity, i.e. a younger age. 
We come back to this problem in the next paragraph. First, we describe how the 
photospheric UV fluxes have been derived. 
We can avoid directly dealing with the uncertain radius by computing 
the photospheric UV flux density from the
$UV - J$ color predicted from the DUSTY model and the observed $J$-band flux density 
($(f_{\rm J})_\lambda$), 
\begin{equation}
(f_{\rm UV,ph})_\lambda = \frac{(F_{\rm UV,DUSTY})_\lambda}{(F_{\rm J,DUSTY})_\lambda} 
\cdot (f_{\rm J})_\lambda 
\label{eq:photo_UV}
\end{equation}
The theoretical $UV$ and $J$-band flux densities, $(F_{\rm UV,DUSTY})_\lambda$ and 
$(F_{\rm J,DUSTY})_\lambda$, 
were obtained by convolving the synthetic spectrum with the respective normalized 
filter transmission curve as described by \cite{Bayo08.1}. 
The photospheric UV flux densities from Eq.~\ref{eq:photo_UV}
were converted to fluxes ($f_{\rm UV,ph}$) 
in the same manner as the observed flux densities by multiplying with the 
effective filter band width (see Sect.~\ref{subsect:act_UV}). 

To test the effect of $\log{g}$ and radius on the photospheric UV emission we have repeated 
the evaluation of Eq.~\ref{eq:photo_UV} for DUSTY models of $\log{g} = 4.5$. This gravity 
corresponds to an age of $\log{t}\,{\rm [Myr]} = 7.5$ according to \cite{Baraffe98.1}.
We find that the photospheric UV flux is smaller and the UV excess
larger by up to a factor of two with respect to our assumption of $\log{g} = 5$. 
However, the age related to this low gravity would imply a pre-main sequence status
for a large fraction of the $10$-pc sample. 
An investigation of the age of individual stars on
the basis of a comparison to model predictions is not the scope of this paper. 
Therefore, we use throughout the remainder of this paper the UV fluxes and chromospheric
excess fluxes derived 
for $\log{g} = 5.0$ but keep in mind that they may be up to a factor two higher in 
individual cases.

Similarly to the UV, the ratio of X-ray flux to bolometric flux is defined as $R_{\rm x}$.  
Here no photospheric flux is considered because the photosphere is
not expected to produce emission in the X-ray band. 

For the H$\alpha$ emission we resort directly to the values computed by 
\cite{Reiners12.0} and \cite{Mohanty03.1}.

\section{Characteristics of activity on M dwarfs}\label{sect:results}

In the following we examine the connection between various activity diagnostics and 
their relation with other stellar parameters such as effective temperature and rotation 
rate. For the UV which is poorly explored so far we provide a comparison to available 
literature data. 
Unless stated otherwise, all our UV measurements refer to the chromospheric excess flux after
subtraction of the photospheric contribution as described in Sect.~\ref{subsect:act_rindex}.

\subsection{Chromospheric vs photospheric UV flux}\label{subsect:results_uv_chrom_photo}

\begin{figure}
\begin{center}
\includegraphics[width=9cm,angle=0]{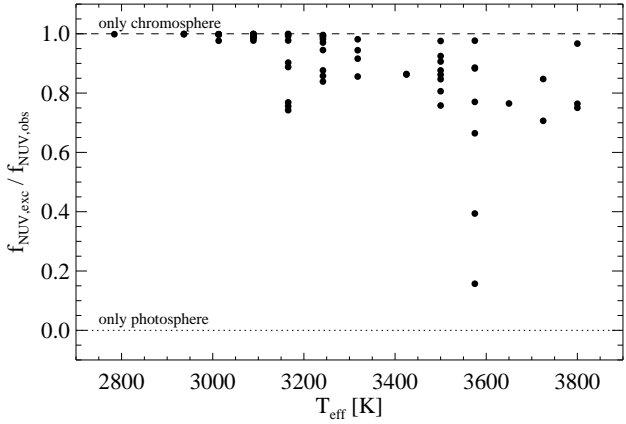}
\caption{Fraction of NUV emission in excess of the photospheric model spectrum.
Horizontal lines define stars for which all NUV emission is from the chromosphere
(dashed) or from the photosphere (dotted).}
\label{fig:uvexcess}
\end{center}
\end{figure}

The chromospheric emission resulting in excess flux with respect to the photospheric 
model is shown in Fig.~\ref{fig:uvexcess} for the NUV band. Here, the excess is normalized
to the observed NUV flux. 
The lower envelope of these data represents the maximum observed contribution of the
photosphere to the NUV emission for a given $T_{\rm eff}$. 
For the majority of stars the dominating contribution to the NUV emission comes 
from the chromosphere. In fact, no star has purely photospheric NUV emission. 
However, for several stars the photosphere contributes $\sim 20-30$\,\% to the NUV emission,
and for the most extreme cases the photosphere produces more than half of the observed NUV flux.
The higher the effective temperature the larger on average the contribution of
the photosphere to the NUV emission. %, i.e. the lower the lower envelope of the
This is not surprising considering that the spectral energy distribution shifts
towards the blue for hotter stars. 
The fact that the data points are distributed below the horizontal line 
representing $100$\,\% chromospheric NUV flux in Fig.~\ref{fig:uvexcess} 
shows that subtracting the photospheric part of the NUV
emission is essential for a correct assessment of the chromospheric emission. 
% 
% OUTPUT FROM   plot_Rindex.pro
%

An analogous analysis in the FUV has shown that in this energy band all
photospheres in our sample are very faint, such that the chromospheric 
excess amounts to $\geq 99$\,\% of the observed emission.

\subsection{High-energy spectral energy distributions}\label{subsect:results_he_seds}

\begin{figure}
\begin{center}
\includegraphics[width=9.0cm,angle=0]{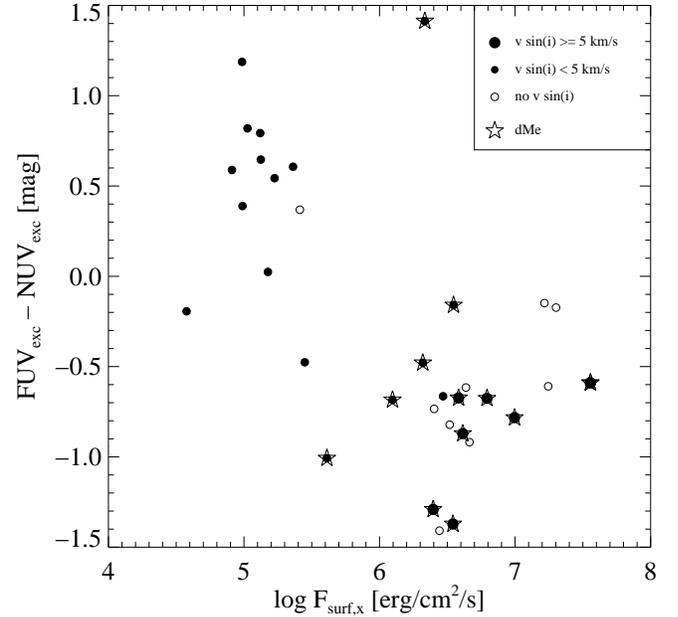}
\caption{`Color' representing FUV vs. NUV excess in function of X-ray surface flux.
H$\alpha$ emitting stars (dMe) are highlighted. Fast and slow rotators and stars
without information on $v \sin{i}$ are distinguished by different plotting symbols.}
\label{fig:he_colors}
\end{center}
\end{figure}

The `spectral energy distribution' of the excess emission can be studied by means
of a UV color representing the excess emission and defined as 
\begin{equation}
FUV_{\rm exc} - NUV_{\rm exc} = -2.5 \cdot \log{(\frac{f_{\rm exc,FUV}}{f_{\rm exc,NUV}})} 
\end{equation}
where $f_{\rm exc,FUV}$ and $f_{\rm exc,NUV}$ are the chromospheric excess fluxes 
%$f_{\rm UV,obs} - f_{\rm UV,ph}$ 
defined in Sect.~\ref{subsect:act_rindex}.
In Fig.~\ref{fig:he_colors} we show this color in dependence of the X-ray surface
flux as a measure of the activity level. Only stars detected in all three energy bands
(X-rays, FUV and NUV) are shown. For highly (X-ray) active stars the FUV excess
dominates the NUV excess, while for weaker X-ray emitters the NUV excess is stronger
than the FUV excess. 

\subsection{Relations between H$\alpha$, UV and X-ray emission}\label{subsect:results_flux_flux}

Various chromospheric lines have been shown for different 
samples of late-type main-sequence stars to be connected by a power-law relation
of the type 
\begin{equation}
\log{F_{\rm 1}} = c_{\rm 1} + c_{\rm 2} \cdot \log{F_{\rm 2}}
\label{eq:flux_flux}
\end{equation}
where
$F_{\rm 1,2}$ are the surface fluxes representing two different activity diagnostics 
and $c_{\rm 1}$ and $c_{\rm 2}$ are coefficients.
Early comparisons of coronal X-ray and chromospheric emission line fluxes in samples
of late-type dwarfs 
have suggested a spectral type dependence of the slopes \citep[e.g.][]{Schrijver87.1, Doyle89.1}.
However, \cite{MartinezArnaiz11.0} found on a much larger sample 
that the flux-flux relations are universal in
spectral type but with a separate branch for the most active stars. \cite{Stelzer12.0}
have for the first time extended these relations into the regime of the ultracool
(late-M) dwarfs.  

\begin{figure*}
\begin{center}
\parbox{18cm}{
\parbox{9cm}{
\includegraphics[width=9.0cm,angle=0]{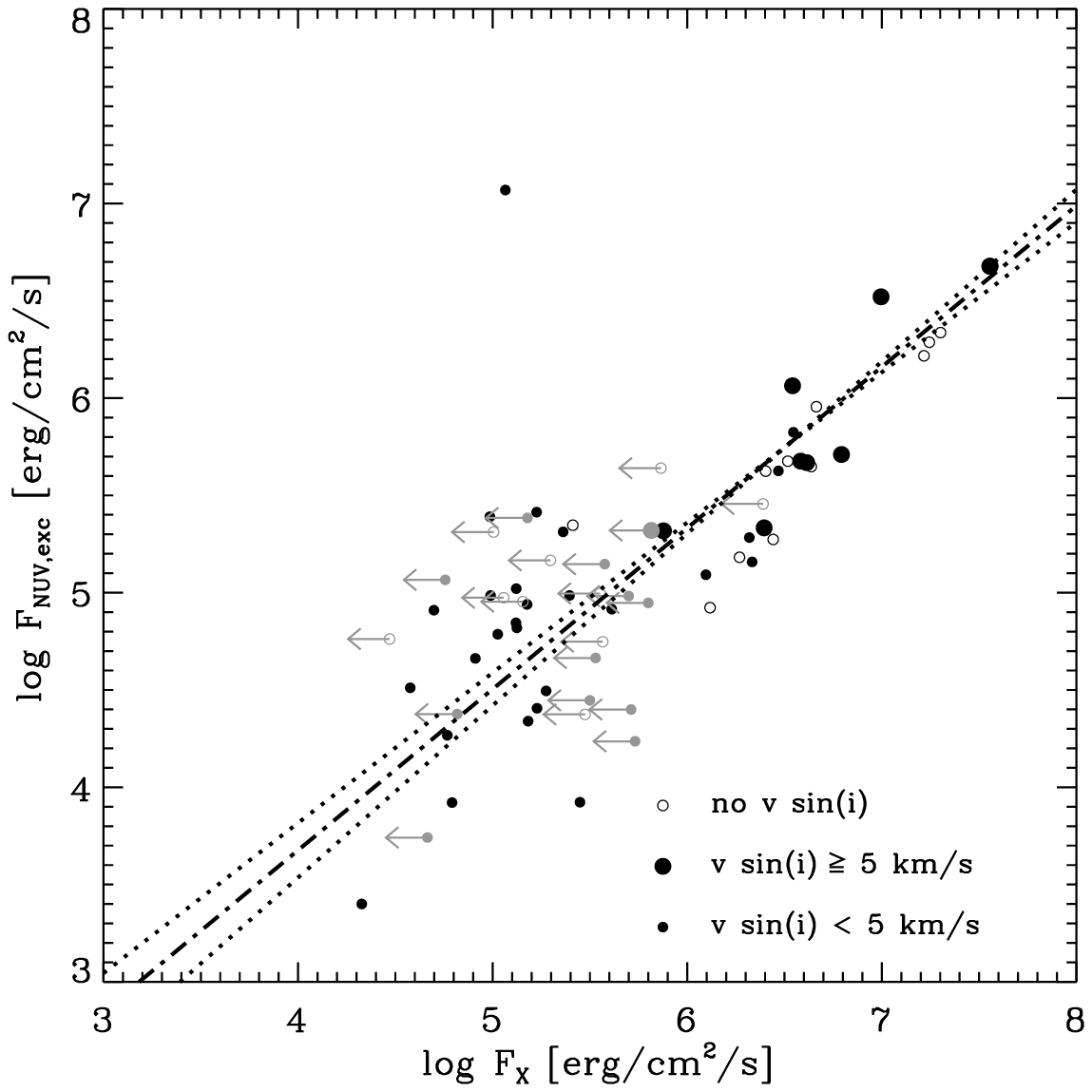}
}
\parbox{9cm}{
\includegraphics[width=9.0cm,angle=0]{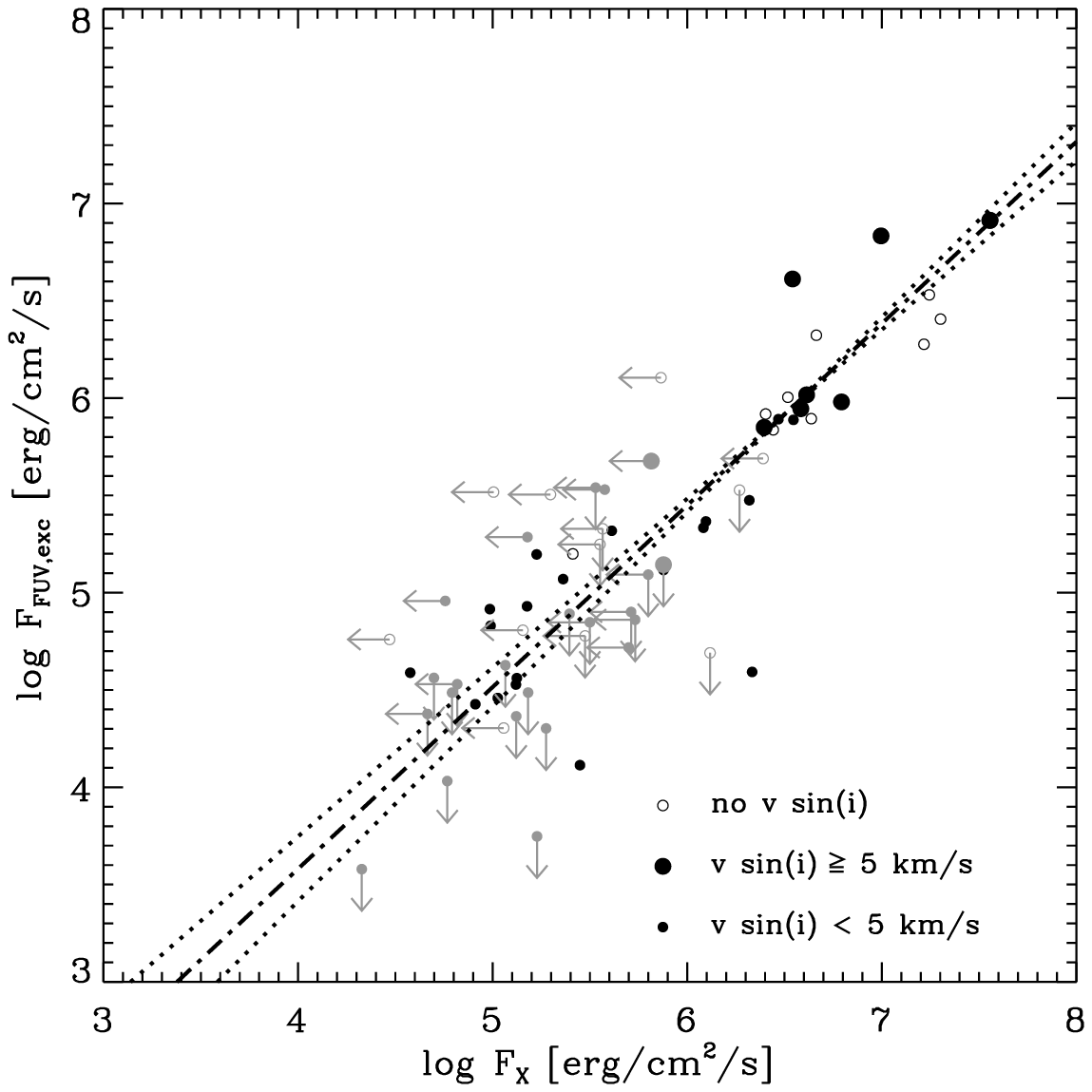}
}
}
\parbox{18cm}{
\parbox{9cm}{
\includegraphics[width=9.0cm,angle=0]{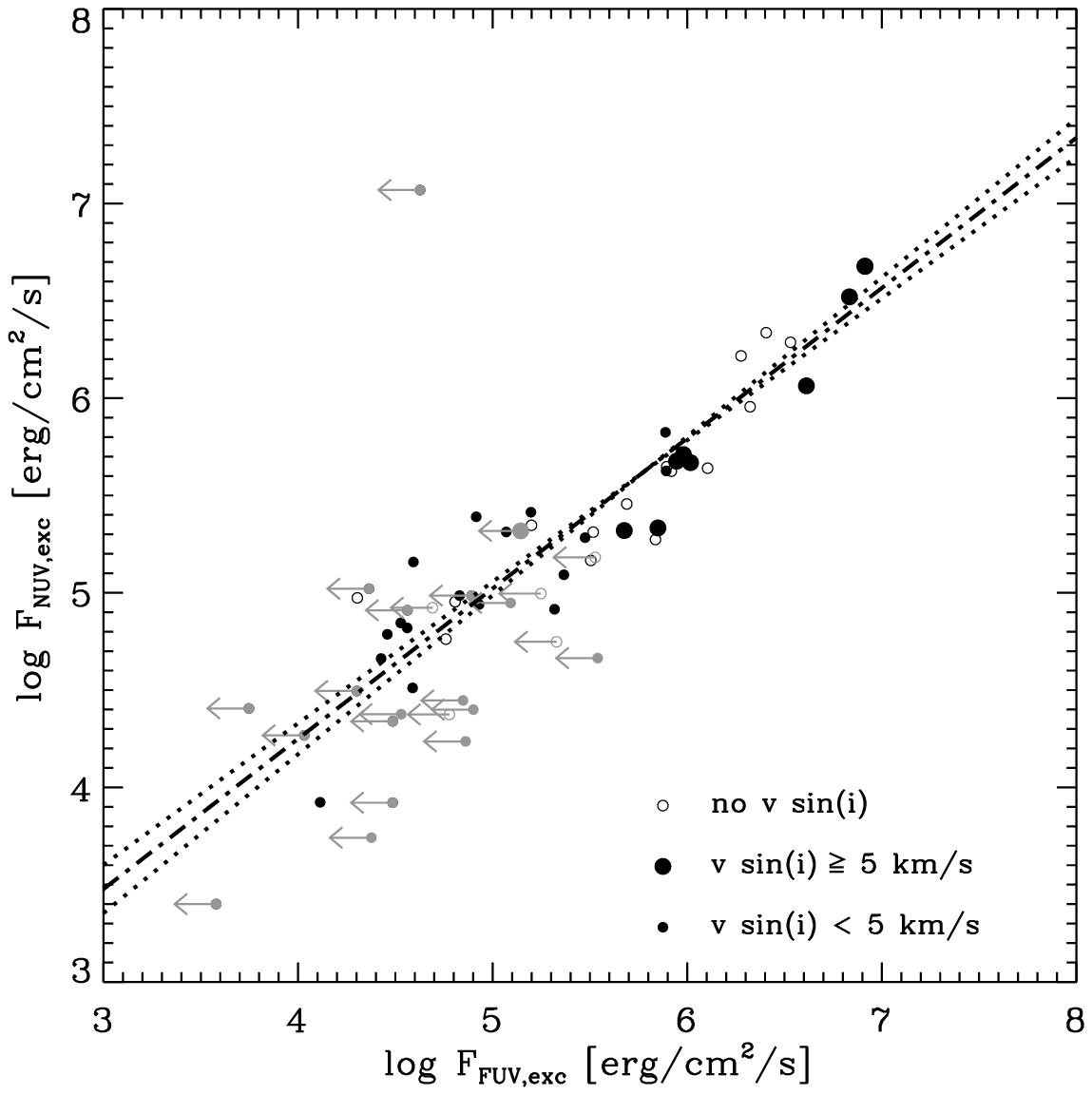}
}
\parbox{9cm}{
\includegraphics[width=9.0cm,angle=0]{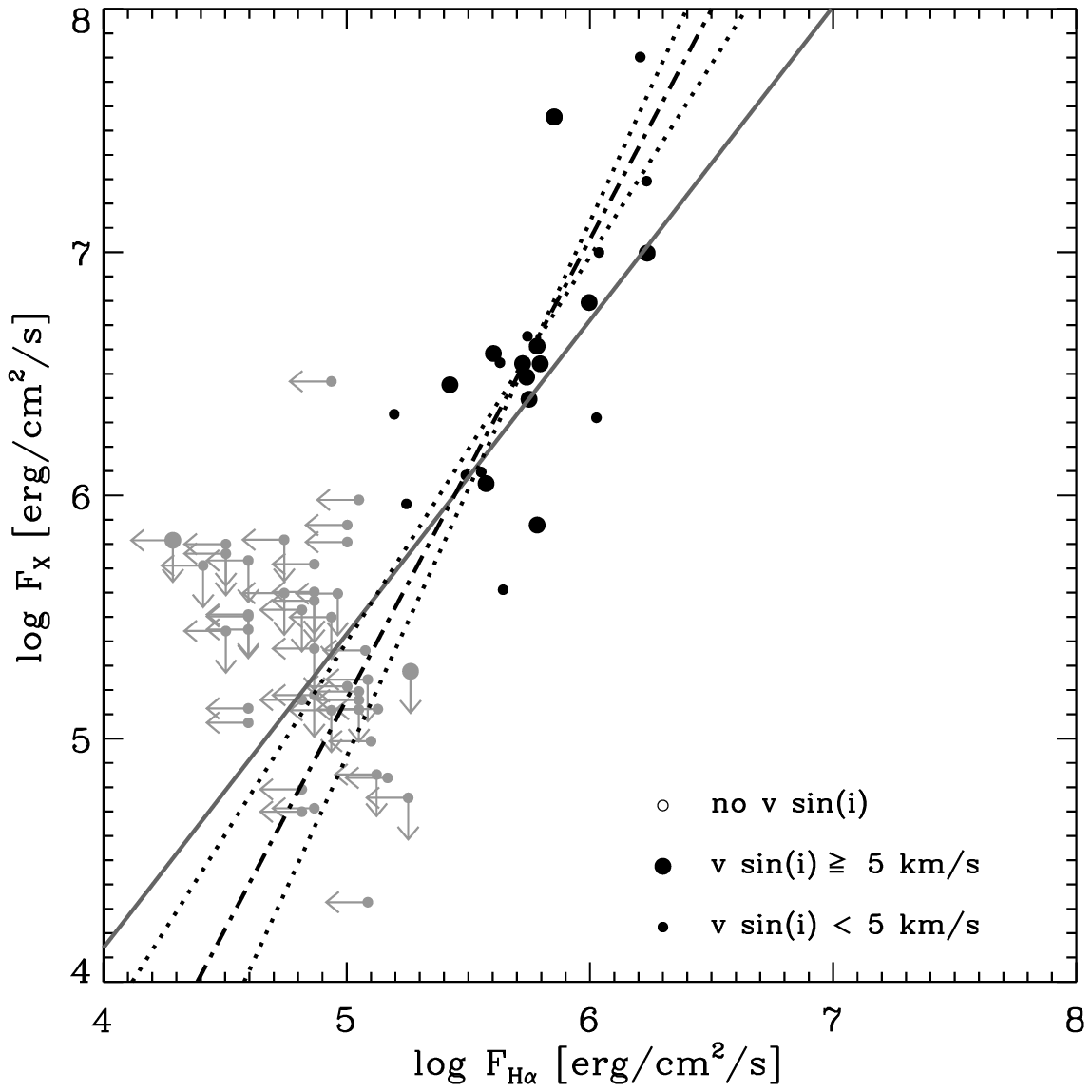}
}
}
\caption{Flux-flux relationships for UV, X-ray and H$\alpha$ emission. The size of the
plotting symbols characterizes the rotation rates. 
Upper limits are drawn as grey arrows. Linear regression fits to the detections (black 
symbols) are shown as dash-dotted lines and their variance as dotted lines. 
The flux-flux relation derived by \protect\cite{Stelzer12.0} for a sample of M dwarfs
including ultracool dwarfs is overplotted as solid line in the bottom right panel.} 
\label{fig:flux_flux}
\end{center}
\end{figure*}

Here, we search for flux-flux relations in the 
broad-band UV and X-ray emission together with H$\alpha$ emission for the $10$-pc sample. 
The surface fluxes
in all activity diagnostics were calculated using the individual radii determined as
described in Sect.~\ref{subsect:act_rindex}. The results are shown in Fig.~\ref{fig:flux_flux}
where the size of the plotting symbols is used to distinguish fast from slow rotators
and different color codes are used for detections and upper limits.

Considering only the detections, the pair of activity diagnostics with the smallest
spread is NUV vs FUV. 
We have performed linear regression fits to different pairs of fluxes considering only
the stars detected in both involved energy bands. As in our previous works, 
the method used is the least squares bisector regression described by \cite{Isobe90.0}. 
The results are overplotted
in Fig.~\ref{fig:flux_flux} and the coefficients of the log-log relations are given in 
Table~\ref{tab:flux_flux}. 
We have omitted one star (Gl\,169.1\,A) from the fit of NUV vs. X-ray emission.
Its very high NUV flux places it orders of magnitude above the bulk of the stars
and may be due to a flare event. 
Note also that the inclusion of the upper limits might modify the relations.

\begin{table}\begin{center}
\caption{Coefficients of double-logarithmic relations of the type shown
in Eq.~\ref{eq:flux_flux} for three different measures for activity (surface flux, 
activity index and luminosity) and four energy bands (H$\alpha$, NUV, FUV, and X-rays).}
\label{tab:flux_flux}
\begin{tabular}{ccccc} \\ \hline
Line 1 & Line 2 & $N_*$ & $c_{\rm 1}$ & $c_{\rm 2}$ \\ \hline
\multicolumn{5}{c}{Surface flux} \\ \hline
X & H$\alpha$ & 24 & $-4.29 \pm 1.77$ & $1.89 \pm 0.31$ \\
FUV & X & 36 & $-0.16 \pm 0.43$ & $0.93 \pm 0.07$ \\
NUV & X & 46 & $0.36 \pm 0.36$ & $0.83 \pm 0.06$ \\
NUV & FUV & 46 & $1.16 \pm 0.26$ & $0.77 \pm 0.04$ \\ \hline
%NUV & X & 45 & $0.05 \pm 0.41$ & $0.87 \pm 0.06$ \\
%NUV & FUV & 46 & $0.81 \pm 0.28$ & $0.83 \pm 0.05$ \\ \hline
\multicolumn{5}{c}{Activity index} \\ \hline
X & H$\alpha$ & 24 & $4.43 \pm 1.23$ & $1.90 \pm 0.31$ \\
FUV & X & 36 & $-0.79 \pm 0.21$ & $0.94 \pm 0.06$ \\
NUV & X & 46 & $-1.39 \pm 0.19$ & $0.81 \pm 0.05$ \\
NUV & FUV & 46 & $-1.13 \pm 0.16$ & $0.76 \pm 0.04$ \\ \hline
%NUV & X & 45 & $-1.27 \pm 0.20$ & $0.86 \pm 0.06$ \\
%NUV & FUV & 46 & $-0.92 \pm 0.17$ & $0.82 \pm 0.04$ \\ \hline
\multicolumn{5}{c}{Luminosity} \\ \hline
X & H$\alpha$ & 24 & $-15.71 \pm 6.27$ & $1.61 \pm 0.23$ \\
FUV & X & 36 & $1.11 \pm 2.26$ & $0.94 \pm 0.08$ \\
NUV & X & 46 & $0.92 \pm 1.70$ & $0.94 \pm 0.06$ \\
NUV & FUV & 46 & $2.58 \pm 1.34$ & $0.90 \pm 0.05$ \\ 
%NUV & X & 45 & $-0.03 \pm 1.98$ & $0.98 \pm 0.07$ \\
%NUV & FUV & 46 & $1.88 \pm 1.45$ & $0.93 \pm 0.05$ \\
\hline
\end{tabular}\end{center}\end{table}

\begin{figure*}
\begin{center}
\parbox{18cm}{
\parbox{6cm}{
\includegraphics[width=6.0cm,angle=0]{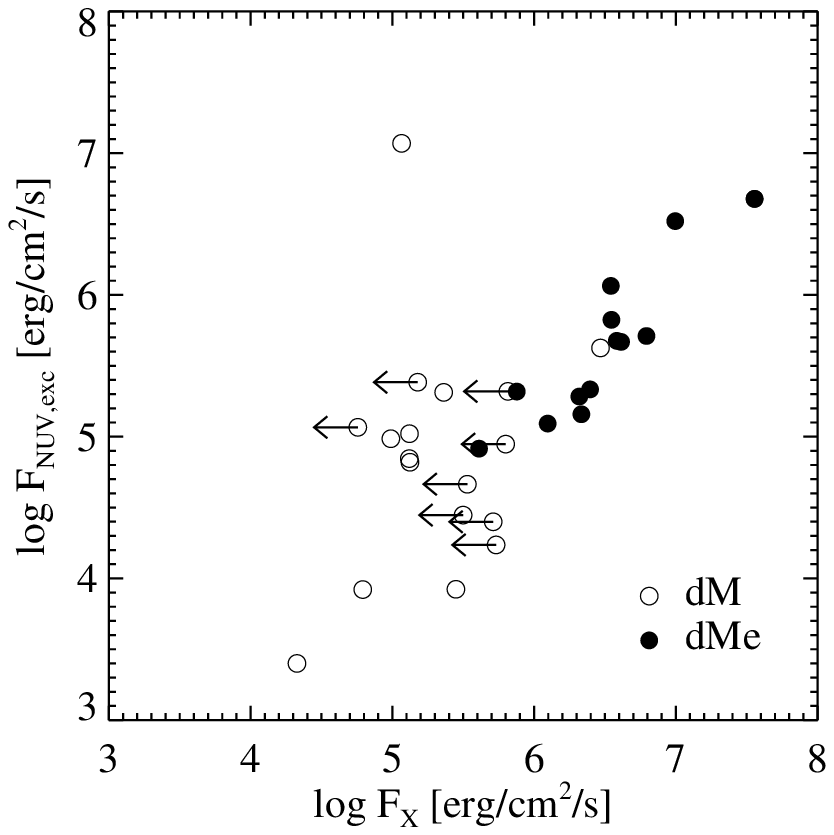}
}
\parbox{6cm}{
\includegraphics[width=6.0cm,angle=0]{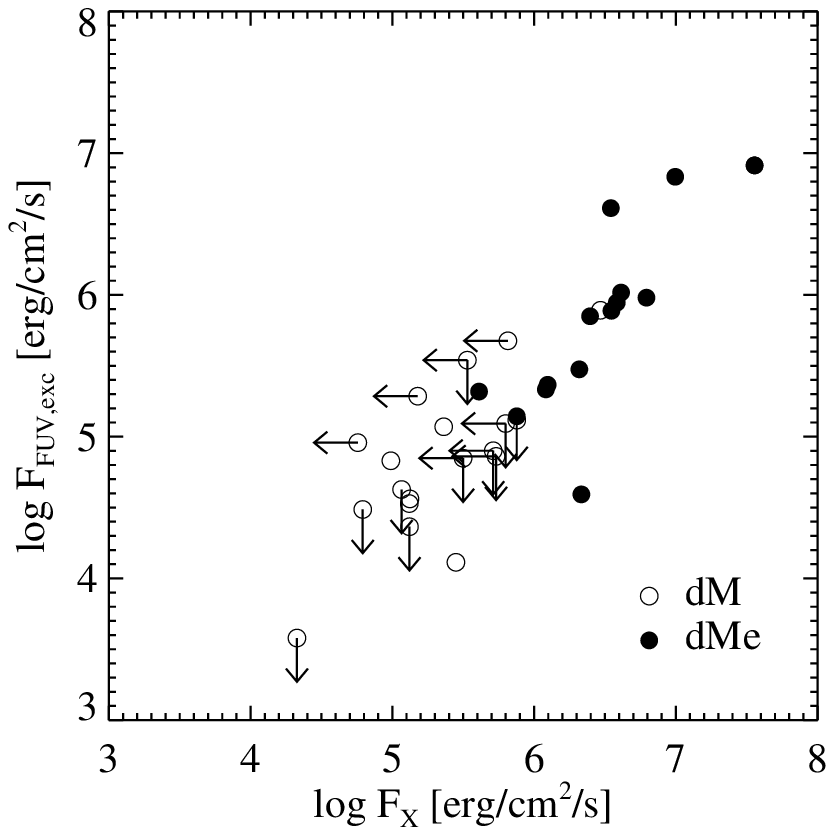}
}
\parbox{6cm}{
\includegraphics[width=6.0cm,angle=0]{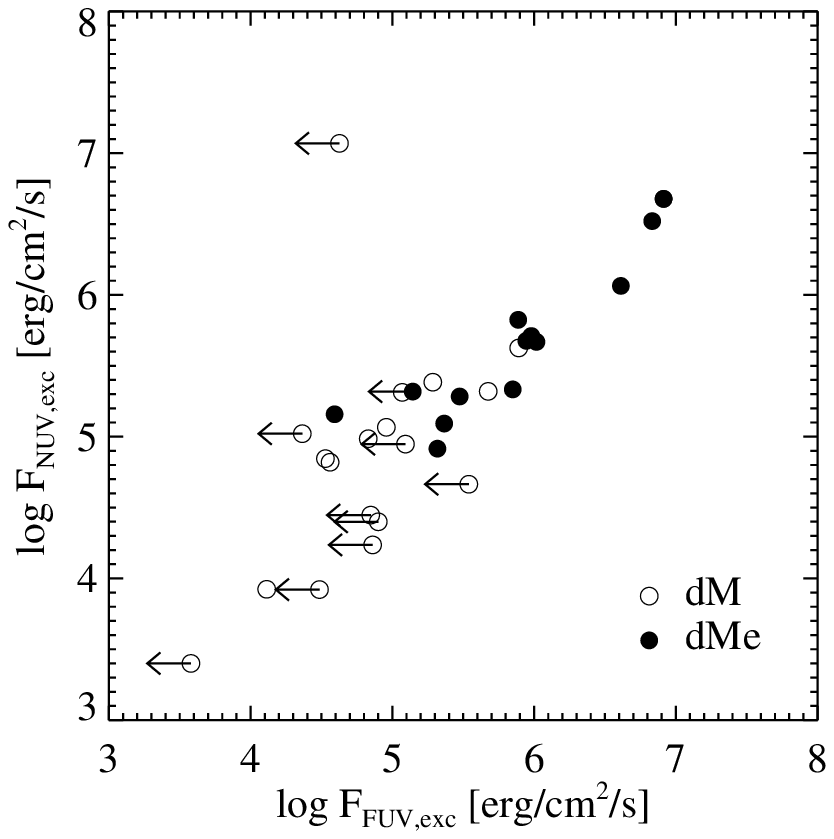}
}
}
\caption{Flux-flux relationships for UV vs. X-ray emission distinguishing H$\alpha$
emitting dMe stars from those without detected H$\alpha$ emission (dM stars).} 
\label{fig:flux_flux_dMdMe}
\end{center}
\end{figure*}

Another representation of the flux-flux relationships is shown in 
Fig.~\ref{fig:flux_flux_dMdMe}.
Here, for the comparison of X-ray and UV fluxes we consider separately H$\alpha$ emitters
(dMe stars) and stars with upper limit to the H$\alpha$ flux (dM stars). 
Fig.~\ref{fig:flux_flux_dMdMe}
shows that the dM stars are generally located on the extension of the flux-flux relation
for dMe stars.

Relations between luminosities or activity indices of two 
spectral bands can be examined analogous to the case of the flux-flux relationships. 
The results are summarized in Table~\ref{tab:flux_flux}.

\subsection{Activity indices vs. spectral type and rotation rate}\label{subsect:results_rindex}

\begin{figure*}
\begin{center}
\parbox{18cm}{
\parbox{9cm}{
\includegraphics[width=9cm,angle=0]{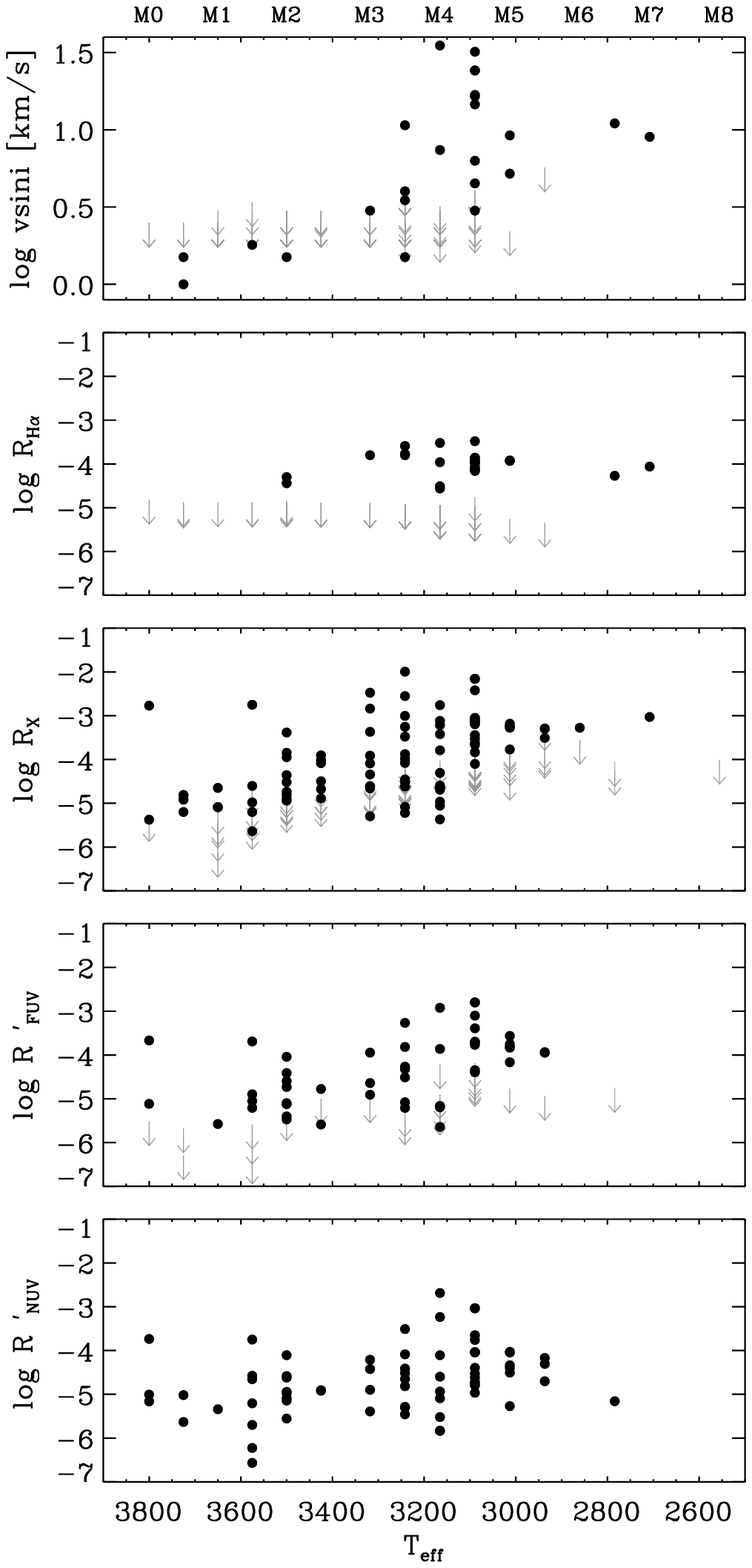}
}
\parbox{9cm}{
\includegraphics[width=9cm,angle=0]{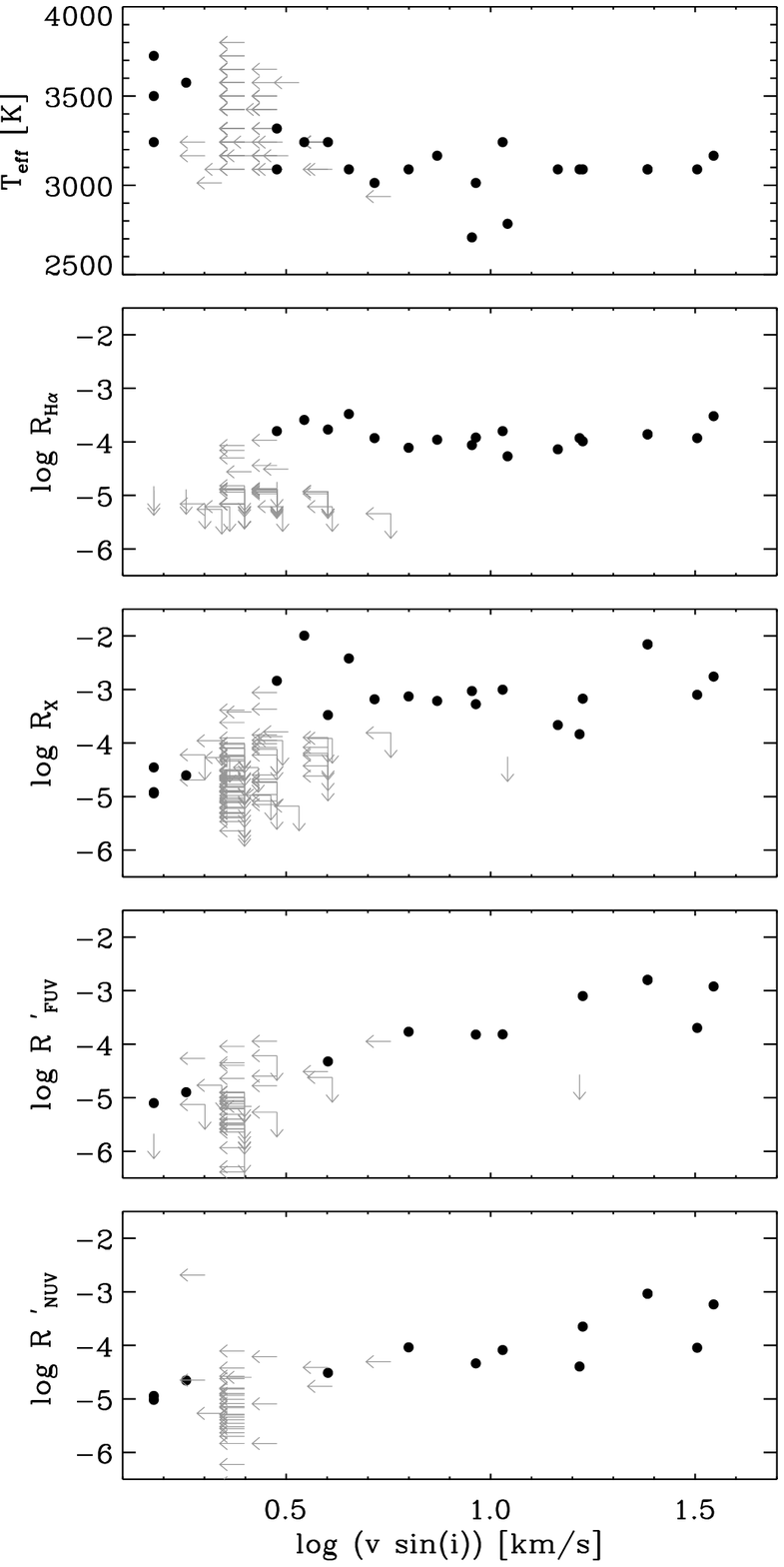}
}
}
\caption{Activity indices in the X-ray, FUV and NUV bands and H$\alpha$ emission 
versus effective temperature (left) and versus rotational velocity (right). 
Upper limits are represented by grey arrows. 
For the FUV and NUV only the excess emission over the photosphere is taken into account. 
The topmost panels show as a reference $\log{v \sin{i}}$ (left) and $T_{\rm eff}$ 
(right), respectively.} 
\label{fig:act_indices}
\end{center}
\end{figure*}

The distribution of the activity indices for all four diagnostics (H$\alpha$, X-rays, 
FUV and NUV) 
are shown in Fig.~\ref{fig:act_indices} as a function of $T_{\rm eff}$ and $v \sin{i}$.
Upper limits are symbolized by grey downward pointing arrows. 
The H$\alpha$ sample is restricted in spectral type by the available published
spectroscopic data. For conclusions on the H$\alpha$ activity of early-M dwarfs we
refer to \cite{Reiners12.0} 
who have studied a much larger sample. We used the H$\alpha$ data here
as comparison to our new data for the X-ray, FUV and NUV emission of the $10$-pc sample. 

For the X-ray band, 
the lower envelope of the activity index increases with decreasing effective temperature. 
The location of the upper limits suggests that this is related to 
the reduced sensitivity in $R_{\rm X}$ for the cooler stars which are
fainter in terms of bolometric flux. 
There is no obvious slope of the lower envelope for the FUV and the NUV. 

A spread of several orders of magnitude 
is observed in all activity indices for a given $T_{\rm eff}$. 
This spread corresponds to a range of rotational velocities. In fact, the spread
in the activity indices is largest at spectral type around M4 where the largest range
of $v \sin{i}$ is observed. 
For each spectral subclass, fast 
rotators show higher activitity levels than slow rotators 
in agreement with the common rotation/activity paradigm. 

There seems to be a transition in the rotation rates between early-M ($\leq$\,M3)
and mid- to late-M ($\geq$\,M4); see also \cite{Reiners12.0}. 
All early-M dwarfs in the $10$\,pc sample have very low or undetectable rotation rate. 
On the contrary, fast rotation is observed in about half of the stars with spectral
type M4 and later.
In terms of $R_{\rm H\alpha}$, and less pronounced in $R_{\rm x}$, 
the fast rotators form a plateau in the right panels of Fig.~\ref{fig:act_indices}, 
i.e. the level of activity does not depend on rotation. This 
`saturation' is ubiquituously observed in the activity indices of
late-type stars, yet there is no consensus on its origin. \cite{Vilhu84.0} attributed
it to a complete coverage of the stellar surface with active regions. 
Alternative explanations are saturation of the power of the 
dynamo \citep{Vilhu87.0}, a redistribution of the radiative losses between emission
lines and continuum \citep{Doyle96.0} 
and changes of magnetic field structure \citep{Jardine99.0}.

The evidence for saturation in $R^\prime_{\rm NUV}$ and $R^\prime_{\rm FUV}$
is less clear than for $R_{\rm H\alpha}$ and $R_{\rm X}$.  
We can determine the mean level of each activity index in the presumed
saturated regime. 
For this analysis we restricted the sample to stars with $v \sin{i} \geq 5$\,km/s. 
Beyond this value H$\alpha$ and X-ray activity indices have reached a plateau
in the right panel of Fig.~\ref{fig:act_indices}.  
The average activity indices and its standard deviation for this subsample of fast 
rotators are given in Table~\ref{tab:sat_level}. The trend for the broad bands
(NUV, FUV, X-rays) goes for increasing mean activity level with increasing 
atmospheric height. However, the small number of stars evaluated leads to a rather
large spread. The H$\alpha$ emission which is produced in the
chromosphere, i.e. relatively low in the atmosphere, is in agreement with this trend as it 
presents the faintest saturation level. 
\begin{table}
\begin{center}
\caption{Average emission level in the activity indices for fast rotating M dwarfs 
($v \sin{i} \geq 5$\,km/s), standard deviation, and number of stars.}
\label{tab:sat_level}
\begin{tabular}{lccc} \hline
Act. index & mean & std.dev & $N_*$ \\ 
\hline
% RESULTS FOR "SPEC"; CORRECTED CF_ROSAT; CORRECTED GALEX BANDWIDTHS
$\log{R_{\rm X}}$          & $-3.18$ & $0.57$ & $15$ \\
$\log{R^\prime_{\rm FUV}}$ & $-3.52$ & $0.58$ & $10$ \\
$\log{R^\prime_{\rm NUV}}$ & $-3.82$ & $0.54$ & $10$ \\
$\log{R_{\rm H\alpha}}$    & $-4.04$ & $0.40$ & $15$ \\ 
\hline
\end{tabular}
\end{center}
\end{table}

Slow rotators ($v \sin{i} \leq 3$\,km/s) cover 
a large range of observed activity levels. Whether activity depends on rotation in 
that regime is difficult to determine because the available sensitivity limits in terms 
of the rotation rate are too high and because the rotation rate is only a lower
limit because of the unknown inclination angle. 

\subsection{Comparison with published UV data of M dwarfs}\label{subsect:results_uv_publ}

We have searched the literature for spectroscopic data in the UV that
would allow us to evaluate the relative contributions of lines and continuum to
the GALEX fluxes. While detailed UV spectroscopy has been presented for a few
individual M dwarfs \citep[e.g.][]{Pagano00.1, Hawley03.0, Hawley07.1}, 
published chromospheric line fluxes are found for very few stars of the $10$-pc sample. 
Among the most notable and best-studied emission features in dwarf stars are 
the chromospheric Mg\,{\sc II}\,h+k doublet at $\lambda\lambda 2796,2803$\,\AA~ 
in the NUV and the 
C\,{\sc IV}$\lambda\lambda 1548,1550$\,\AA~ doublet from the transition region in the FUV. 
\citet[][henceforth MD89]{Mathioudakis89.1} have listed Mg\,{\sc II} surface fluxes 
($F_{\rm s, MgII}$) for K-M dwarf stars based on IUE spectra. Their sample has $18$ stars 
in common with ours. 
Further fluxes for FUV and NUV continuum as well as some important UV emission lines 
from M dwarfs are presented by 
\citet[][henceforth FFL13]{France12.2} 
%et al., ApJ in press (arXiv:1212.4833; henceforth FFL12) 
in their study of HST 
COS and STIS spectra of six exoplanet host stars. Two of their stars are in common with 
the $10$-pc sample.
The UV emission of one of them, GJ\,876, has been examined in detail in another dedicated
study by \cite{France12.0}. In the following we compare the fluxes given in those works to 
our GALEX measurements. 

As we aimed at a homogeneous data analysis we could not simply compare the surface line 
fluxes given by MD89
to the GALEX fluxes but had to reconvert them to observed fluxes ($f_{\rm obs, Mg\,II}$) 
following the inverse of the procedure that was applied by MD89. 
These authors have computed the ratio $F_{\rm s, MgII}/f_{\rm obs, Mg II}$ 
using a relation that involves the $V$ band magnitude, the $R-I$ color, the
effective temperature and a bolometric correction ($B.C.$); see \cite{Oranje82.1} for details. 
They did not list the photometry they used. To be as
close as possible to their historic measurements 
we have compiled $VRI$ magnitudes from CNS\,3 catalog. Again for
reasons of consistency with MD89, 
we extracted $T_{\rm eff}$ and $B.C.$ from \cite{Johnson66.1}.
The observed Mg\,{\sc II} fluxes recovered this way were then transformed back into
surface fluxes using the distances from LG11 and the stellar radii derived in 
Sect.~\ref{subsect:act_rindex}. 
The observed fluxes listed by FFL13 were transformed into surface fluxes in the same way. 

\begin{figure}
\begin{center}
\includegraphics[width=8.5cm,angle=0]{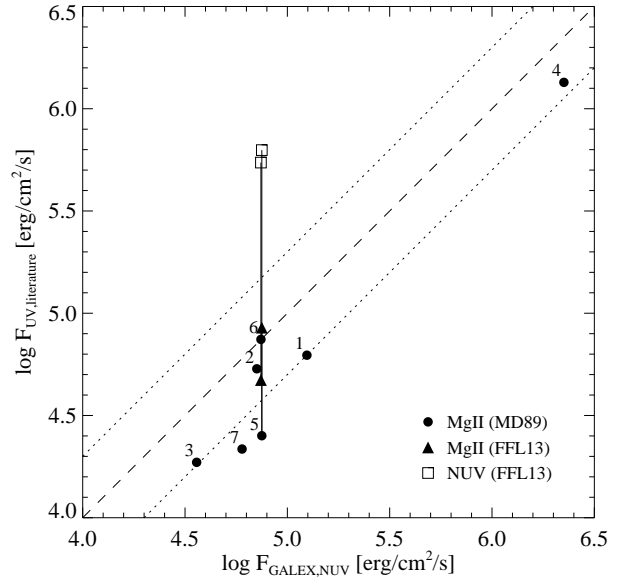}
\caption{Flux-flux relation for 
NUV measurements from the literature and GALEX/NUV broad band. The literature data
comprises Mg\,{\sc II}$\lambda\,2800$\,\AA~ line emission 
published by MD89 (circles) and by 
FFL13 (triangles) and broad band emission in $1150-1790$\,\AA~ 
excluding Ly$\alpha$ (FFL13).  
The Mg\,{\sc II}$\lambda\,2800$\,\AA~ data points from MD89 are labeled with numbers
corresponding to the following stars: 1 -- Gl\,83.1, 2 -- Gl\,411, 3 -- Gl\,412\,A, 4 --
Gl\,803, 5 -- Gl\,832, 6 -- Gl\,876, 7 -- Gl\,908. Two of them (Gl\,832 and Gl\,876)
have data from FFL13. All literature data for these two stars are
connected with a vertical line. The dashed line denotes the 1:1 relation and the 
dotted lines represent a factor two difference between the two measurements. 
}
\label{fig:flux_flux_MgII}
\end{center}
\end{figure}

We plot the flux-flux relations for the GALEX/NUV band and various NUV measurements
from the literature in Fig.~\ref{fig:flux_flux_MgII}. Filled plotting symbols are for 
Mg\,{\sc II} line fluxes and open symbols for broad band NUV fluxes from HST STIS. 
Objects with Mg\,{\sc II} line fluxes from MD89 and from FFL13 are shown with 
different plotting symbols. Several literature measurements for a given star are connected by
a vertical line. 
To make the GALEX fluxes comparable to the literature values we plotted here the observed 
fluxes that include also the photospheric contribution. 
Fig.~\ref{fig:flux_flux_MgII} shows that the two available total NUV fluxes from HST STIS 
are nearly a factor ten higher than those from GALEX.
The Mg\,{\sc II} fluxes are on average roughly a factor two lower than the GALEX/NUV fluxes.  

Analogous to the investigation of the NUV line and broad-band emission, we have
compared the GALEX/FUV to the C\,{\sc IV} and to the broad band FUV emission from 
HST COS for the two stars from the $10$-pc sample with data in FFL13. 
We find that the two stars behave very differently. Similar to the NUV, 
for Gl\,876 the HST/FUV flux
is much larger than the GALEX/FUV flux ($F_{\rm FUV,HST}/F_{\rm FUV,GALEX} \approx 6.5$) 
while the C\,{\sc IV} line flux is similar to the GALEX/FUV flux
($F_{\rm C\,IV}/F_{\rm FUV,GALEX} \approx 0.8$). 
On the contrary, for Gl\,832 the broad band fluxes of the two instruments are within a
factor two ($F_{\rm FUV,HST}/F_{\rm FUV,GALEX} \approx 0.5$) and the C\,{\sc IV} flux is much
smaller than the GALEX/FUV flux ($F_{\rm C\,IV}/F_{\rm FUV,GALEX} \approx 0.04$). 

The comparison of the broad band fluxes must take
into account the different wavelength ranges of the instruments. In the case of the 
NUV this is $2300-3050$\,\AA~ for STIS and $1771-2831$\,\AA~ for the GALEX filter. 
In particular, we recall that the bulk of Mg\,{\sc II} is not being captured by the 
GALEX NUV band which has very low transmission near the cutoff of the filter. 
However, Mg\,{\sc II} makes only about 10\,\% of the total STIS NUV emission.
Therefore, the differences in the NUV broad band fluxes of the two instruments are likely
due to variability. 
Similarly, variability may be responsible for the differences between the FUV fluxes 
observed with GALEX and with the two HST instruments.  
In fact, both stars from FFL13 showed evidence of FUV flaring activity during the HST 
observations while no information is given by FFL13 on the NUV variability. We note that
the HST STIS and COS observations of a given star are not simultaneous. 
The overall impact of the flares onto the FUV emission of the stars could not be quantified
because no lightcurves are presented by \cite{France12.0} and FFL13 
for C\,IV nor for the broad band emission.

\subsection{Age evolution}\label{subsect:results_age}

With the aim to study the age evolution of the activity of M dwarfs 
we have compiled a catalog of all known M-type members of the TW\,Hya association 
(TWA). The TWA represents an excellent comparison sample of pre-main sequence M dwarfs 
because at its age of $10$\,Myr most association members have terminated the
accretion process such that the H$\alpha$, UV and X-ray emission can be
clearly ascribed to magnetic activity.

\subsubsection{TWA sample and analysis}\label{subsubsect:results_age_sample}

Our TWA catalog comprises all stars considered to be association members by
\cite{Mamajek05.0} and $3$ further members discovered since by 
\cite{Looper07.1, Looper10.0, Looper10.1, Shkolnik11.0}.
The X-ray and UV data were extracted in the same way as those of the field M dwarfs 
(see Sects.~\ref{subsubsect:sample_xrays} and~\ref{subsubsect:sample_uv}), 
except that we have consulted only
the RASS BSC and FSC catalogs and no X-ray catalog of pointed observations. 

A significant fraction of TWA members are close binaries that are not resolved
in the RASS and with GALEX. For these cases 
we have distributed the observed X-ray and UV flux equally on both components, 
unless for TWA11 where the primary is 
an A-type star and the X-ray emission is likely produced by the secondary M-type component.
Limiting the sample to M spectral types and treating the binaries as
described, the TWA sample comprises $23$ objects. Sixteen of these are identified
with a RASS/BSC source and one with a RASS/FSC source, while $10$ are 
detected with GALEX. 

For the X-ray count-to-flux conversion a thermal model with $kT = 1$\,keV and 
$N_{\rm H} = 10^{20}\,{\rm cm^{-2}}$ is assumed.  The larger temperature 
with respect to the $10$-pc sample reflects the younger age and the larger column
density the larger distance of TWA. Individual distances were used for the
conversion of fluxes to luminosities.  

The spectral types were converted to effective temperature using the scale of
\cite{Luhman99.1} that was devised for pre-main sequence stars with gravities
inbetween those of dwarfs and of giants. We estimated the bolometric luminosity and
stellar radii of the TWA stars from the \cite{Baraffe98.1} models assuming an age 
of $10$\,Myr. 
Following the procedure described in Sect.~\ref{subsect:act_rindex}, 
we have computed and subtracted the photospheric contribution to the GALEX fluxes for 
the TWA sample. 
Consistent with the younger age of TWA with respect to the $10$-pc sample 
we use the DUSTY models with $\log{g}=4.0$. 
As in the case of the $10$-pc sample, the FUV emission can be entirely
ascribed to the chromospheric excess. Less than $5$\,\% of the NUV emission of TWA
M dwarfs comes from the photosphere.

\subsubsection{Results}\label{subsubsect:results_age_results}

\begin{figure*}
\begin{center}
\parbox{18cm}{
\parbox{6cm}{
\includegraphics[width=6cm,angle=0]{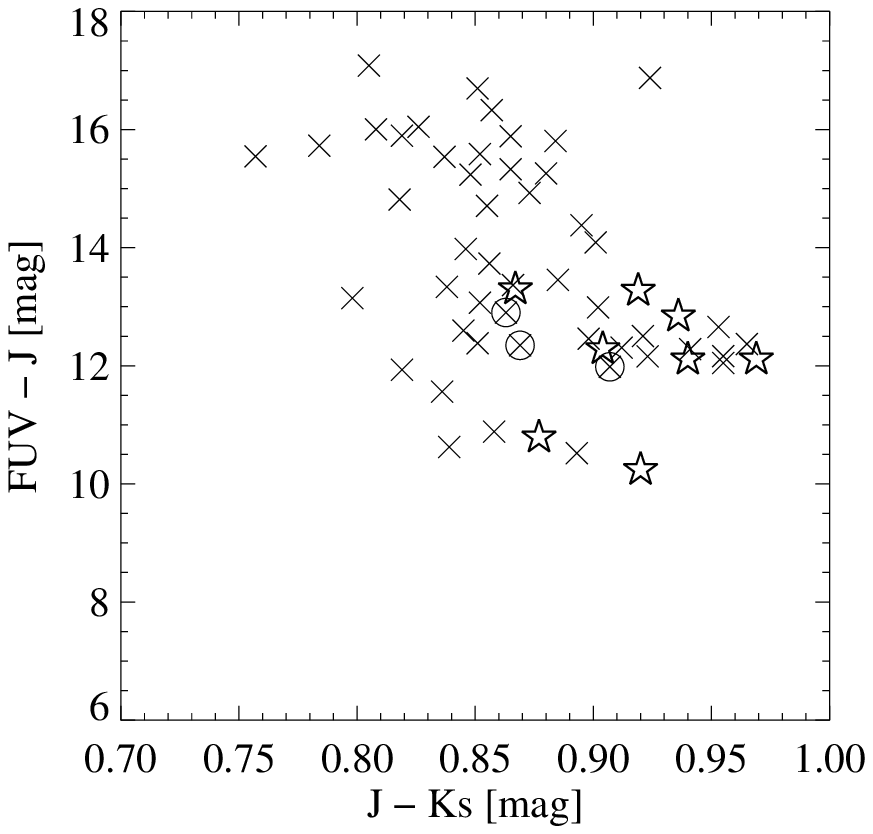}
}
\parbox{6cm}{
\includegraphics[width=6cm,angle=0]{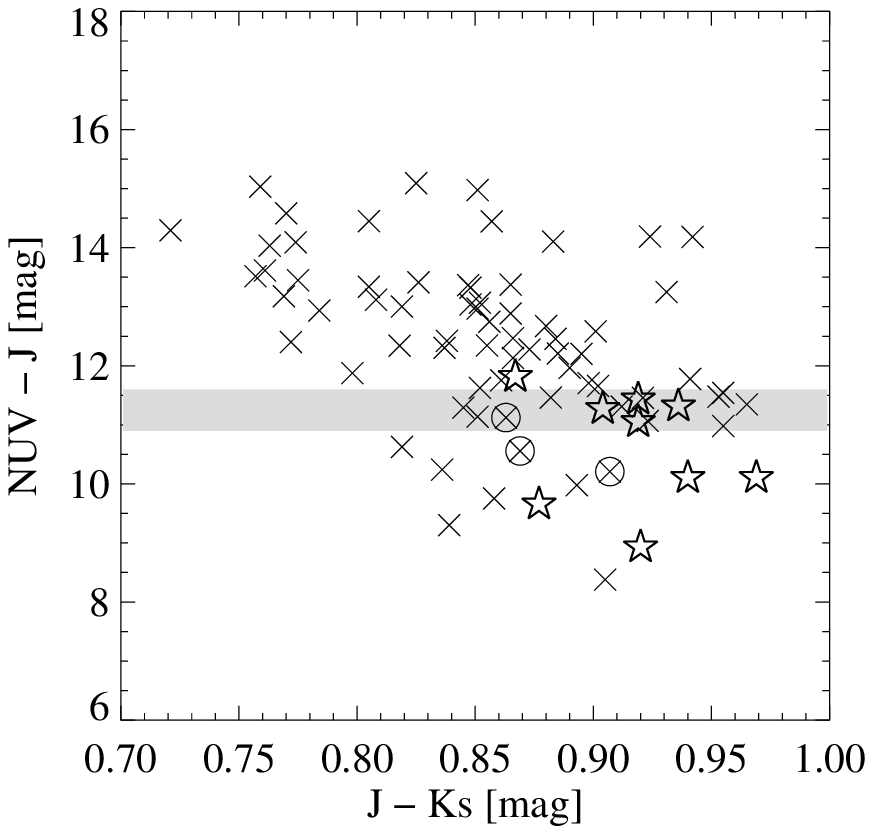}
}
\parbox{6cm}{
\includegraphics[width=6cm,angle=0]{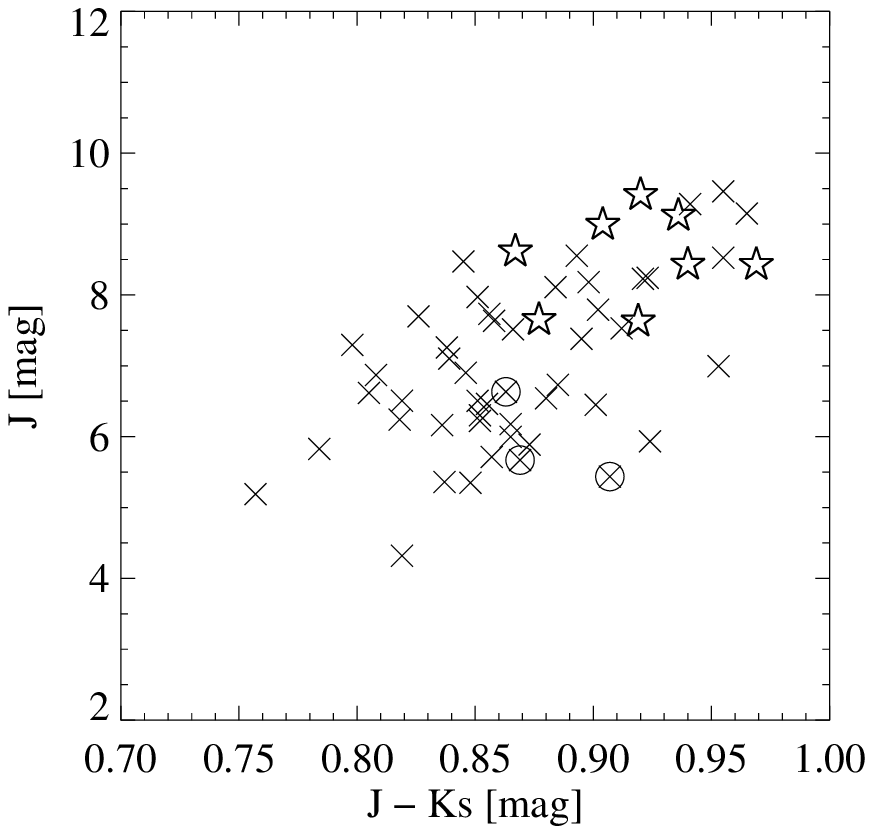}
}
}
\caption{$UV - J$ vs $J - K_{\rm s}$ diagrams and $J$ vs $J - K_{\rm s}$ diagram 
for the $10$-pc sample (crosses) and the TWA sample (star symbols). Crosses surrounded by 
annuli highlight three young or exceptionally active stars 
(AU\,Mic, FK\,Aqr and Gl\,569\,AB) in the $10$-pc sample.  
 The grey-shaded area in the middle panel represents the range of $NUV - J$ color 
observed for Hyades stars with $J > 0.7$ from \protect\cite{Findeisen11.0}.} 
\label{fig:UV_ccds}
\end{center}
\end{figure*}

\cite{Findeisen11.0} have examined the age evolution of GALEX UV magnitude for a sample of 
G stars from various clusters and associations covering $\sim 10$ to $\sim 600$\,Myrs.
In Fig.~\ref{fig:UV_ccds} we show a $UV - J$ versus $J - K$ diagram for the $10$-pc sample
and the TWA sample to be compared to Fig.7 of \cite{Findeisen11.0}. 
We confirm the result of these authors that younger stars have bluer 
$UV - J$ color at the same $J - K$ and we add more detail to it. In particular, 
as can be seen in Fig.~\ref{fig:UV_ccds}, for the same range of $J - K$ 
the majority of M dwarfs from the $10$-pc sample have much redder $NUV - J$ color 
than the oldest sample (Hyades cluster) analysed by \cite{Findeisen11.0}. 
This is likely to be attributed to the older age of the $10$-pc sample with
respect to the Hyades.  
As far as the FUV is concerned, as stated by \cite{Findeisen11.0} their data 
is dominated by non-detections
and we could not derive the $FUV - J$ level of the Hyades from their Fig.7. 
Our Fig.~\ref{fig:UV_ccds} also shows that the anti-correlation in $UV - J$ vs $J - K$ 
that we see in the $10$-pc sample 
is a result of the increasing $J$ band magnitude within the M dwarf sequence. 

\begin{figure*}
\begin{center}
\parbox{18cm}{
\parbox{6cm}{
\includegraphics[width=6.0cm,angle=0]{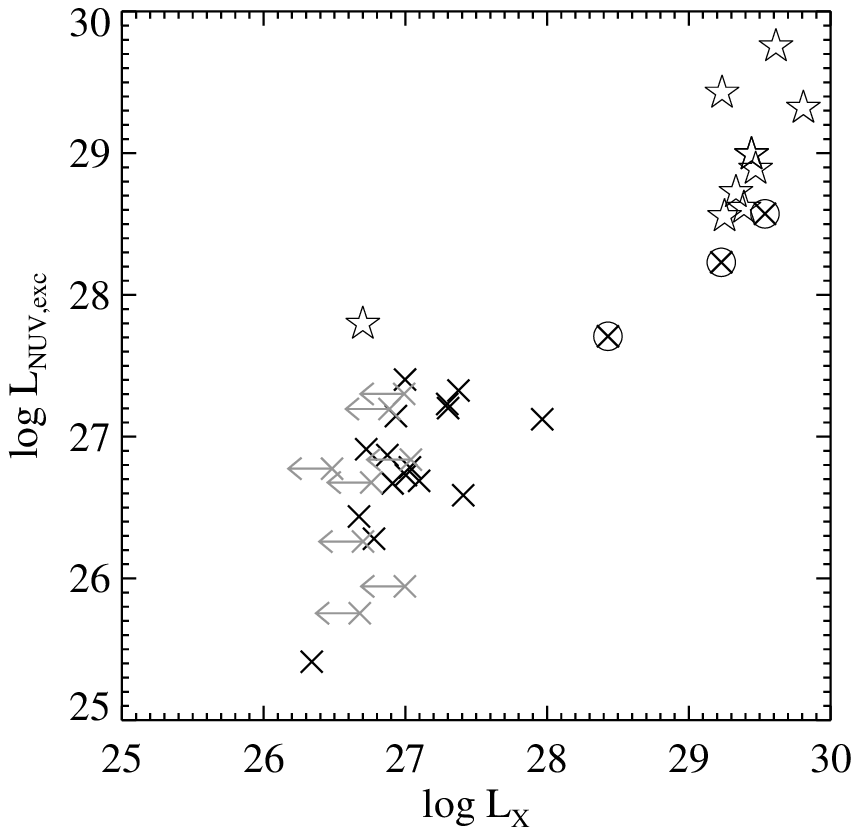}
}
\parbox{6cm}{
\includegraphics[width=6.0cm,angle=0]{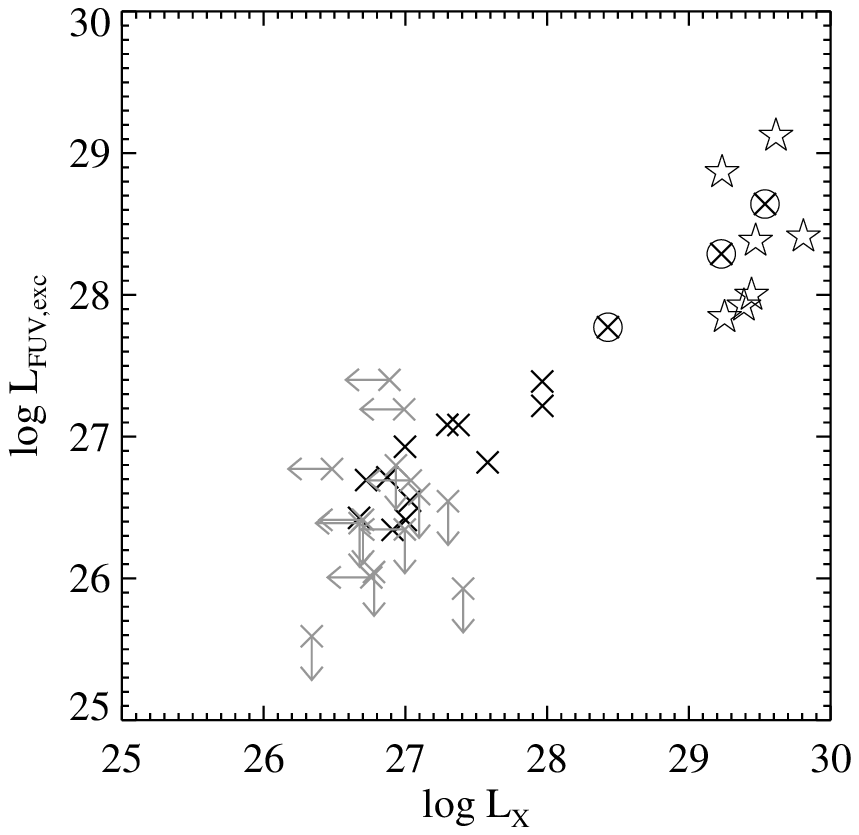}
}
\parbox{6cm}{
\includegraphics[width=6.0cm,angle=0]{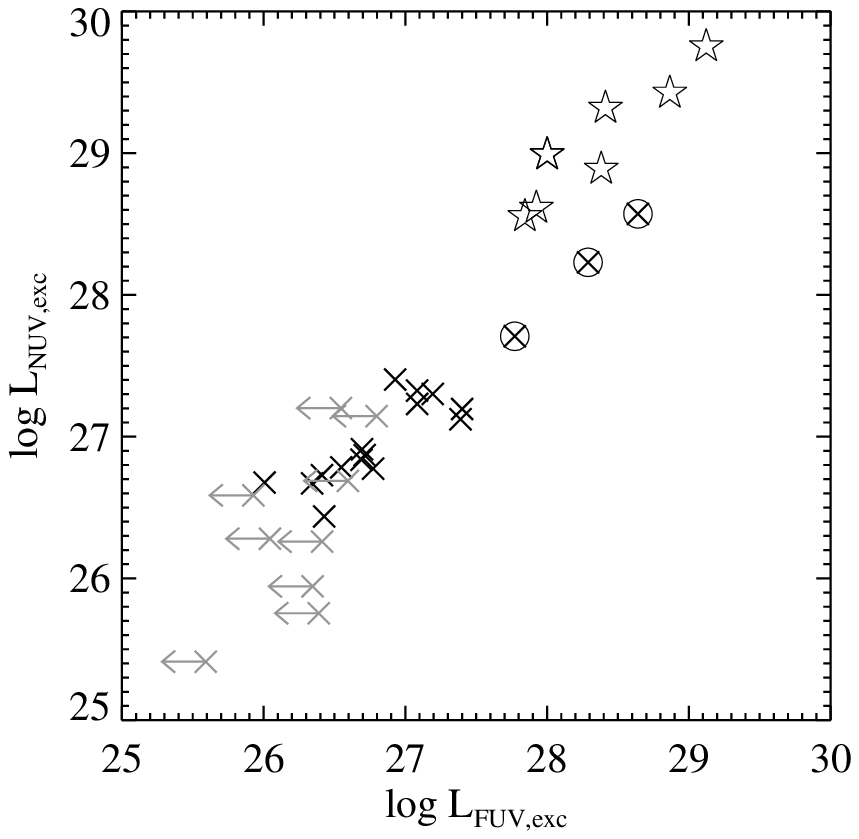}
}
}
\parbox{18cm}{
\parbox{6cm}{
\includegraphics[width=6.0cm,angle=0]{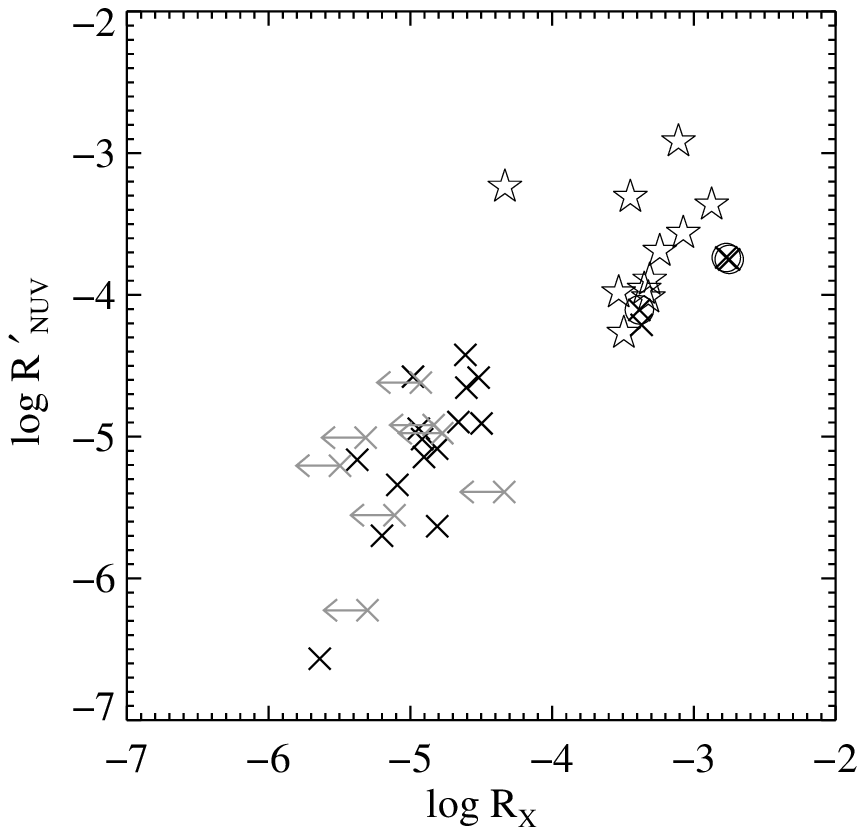}
}
\parbox{6cm}{
\includegraphics[width=6.0cm,angle=0]{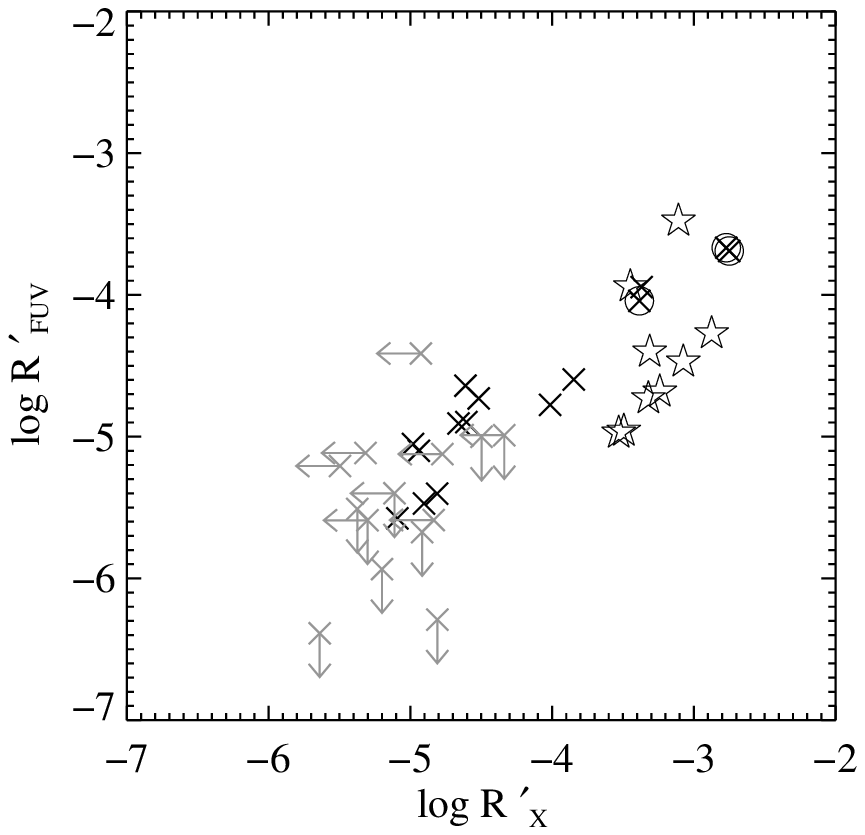}
}
\parbox{6cm}{
\includegraphics[width=6.0cm,angle=0]{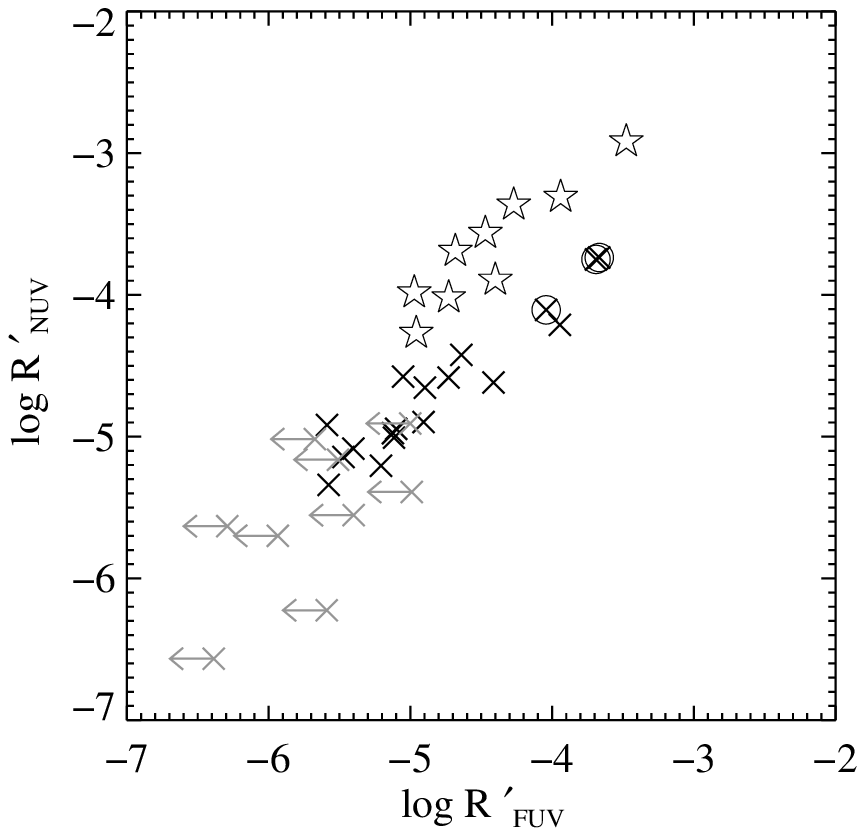}
}
}
\caption{X-ray vs UV luminosity and activity index for M0...M3 stars from 
the $10$-pc sample (crosses)
and stars in the same spectral type range in TWA (star symbols). The three young stars
AU\,Mic, FK\,Aqr and Gl\,569\,AB in the $10$-pc sample are highlighted with annuli 
surrounding the plotting symbol. Upper limits are shown in grey.}
\label{fig:lx_luv_age}
\end{center}
\end{figure*}

In Fig.~\ref{fig:lx_luv_age} the luminosities and activity indices of the 
young M stars in TWA are compared to those of the M0...M3 stars of the 
$10$-pc sample. The restriction to early-M spectral types was applied to the $10$-pc 
sample because all but one of the UV detected TWA stars have early M spectral types; cf.
Fig.~\ref{fig:UV_ccds}. The
only late-M TWA member is TWA-28, detected in NUV and X-rays and clearly offset from the
other TWA stars towards weaker activity in Fig.~\ref{fig:lx_luv_age}. 

Evidently, for the same spectral type range 
the TWA and the $10$-pc sample form two distinct groups in terms of UV and X-ray
luminosities with TWA stars being brighter in all three energy bands. Only the three most 
active stars from the $10$-pc sample (AU\,Mic, FK\,Aqr and Gl\,569\,AB; highlighted by
circles in Figs.~\ref{fig:UV_ccds} and~\ref{fig:lx_luv_age}) are located in 
the same area as TWA suggesting that these are young and/or extremely active stars. 
In fact, AU\,Mic is a debris disk system and a known member of the $12$\,Myr old $\beta$\,Pic 
moving group \citep{Kalas04.0, Zuckerman01.2},
FK\,Aqr (Gl\,867\,A) has long been known as a flare stare \citep{Byrne79.0}, and
Gl\,569\,AB is a multiple system of only $\sim 100$\,Myr \citep{Simon06.1}. 
Our assumption of $\log{g}=5$ is probably wrong for these stars but this 
does not affect the UV excess fluxes because their photospheric contribution is negligibly 
small ($\sim 1$\,\%). 

The differences between the activity levels of TWA and the $10$-pc sample 
persists if the activity indices are considered instead of the luminosities 
(Fig.~\ref{fig:lx_luv_age} bottom) 
although the differences are smaller due to the drop of bolometric luminosity across time.
%{\it Is there any accretors among the TWA stars with UV plus X-ray data?}
%

\begin{figure}
\begin{center}
\includegraphics[width=9cm,angle=0]{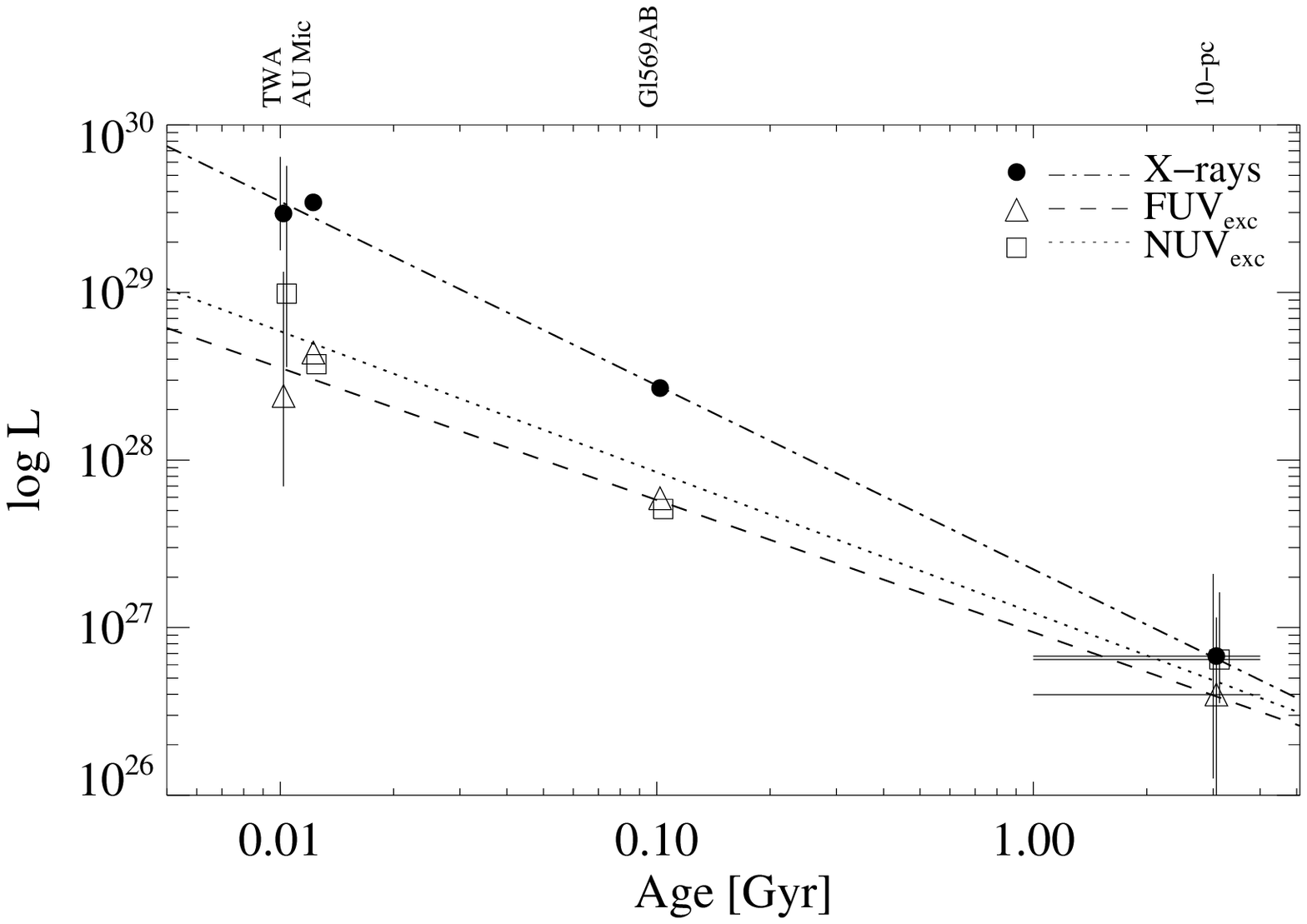}
\caption{Age evolution of UV and X-ray luminosity for M0-M3 stars from the TWA and the
$10$-pc sample. The lines represent regression fits to the three activity diagnostics.}
\label{fig:logl_age}
\end{center}
\end{figure}

The age evolution of the luminosities in the UV and X-ray band for M0-M3 stars
is displayed in Fig.~\ref{fig:logl_age}. The vertical bars for the TWA and the $10$-pc 
sample represent the full spread of observed values. We have omitted the only accretor
in the sample of X-ray and GALEX detected TWA members. Linear regression fits in log-log
form to the four data points are shown for each of the three energy bands. We have 
derived slopes of $\beta_{\rm X} = -1.10 \pm 0.02$, $\beta_{\rm NUV} = -0.79 \pm 0.05$
and $\beta_{\rm NUV} = -0.84 \pm 0.08$. 
Similar plots for the surface flux and activity index vs age show larger scatter of the data
points especially within the TWA sample. This may be related to differences in radii
within the TWA group that are not taken into account in our analysis. 
%
% OUTPUT FROM   plot_act_vs_age.pro
%

\section{Summary and conclusions}\label{sect:discussion}

H$\alpha$, UV and X-ray emission are complementary indicators of magnetic activity
probing the chromosphere, transition region and corona of late-type stars. 
We have quantified the emissions in these wavebands for a well-defined sample of 
nearby M dwarfs 
capitalizing on recent comprehensive catalogs for H$\alpha$ emission and -- partly 
unexplored -- archival data from {\em GALEX}, {\em ROSAT} and {\em XMM-Newton}.

We present a catalog of H$\alpha$, NUV, FUV and X-ray fluxes for $159$ M dwarfs within
$10$\,pc of the Sun, together with the stellar parameters used to derive the fractional
flux in each of the four bands with respect to the bolometric flux, the so-called activity
indices. This sample has an estimated completeness of $\sim 90$\,\%. 

We have subtracted the photospheric contribution to the UV fluxes of all stars with
help of synthetic model spectra. A result of this analysis is that there are
no M dwarfs in 
the $10$-pc sample without a chromospheric contribution to the NUV and FUV emission.
While generally the emission in the GALEX bands is dominated by the chromosphere 
there is a trend towards increasing photospheric contribution for earlier M spectral types
reflecting the shift of the stellar spectral energy distribution towards bluer wavelengths. 
The UV fluxes used throughout this paper for comparison to the H$\alpha$ and X-ray emission
regard only the chromospheric excess remaining after the photospheric subtraction.

As a result of the limitations of the available instrumentation, 
up to the present day UV spectra are available for very few M dwarfs
(see references in Sect.~\ref{sect:intro}).
In particular, the faint continuum has been difficult to assess.  
We found from a comparison of historical records of the Mg\,{\sc II}\,$\lambda 2800$\,\AA~
doublet with GALEX/NUV fluxes that this line 
can contribute about one-third to the NUV emission of M dwarfs. 
% as the whole GALEX band which is probably dominated by continuum flux.  

We have shown that with increasing activity level (measured by the X-ray flux) the relative
contribution of UV emission shifts from the NUV to the FUV. 
Considering that stronger X-ray emitters have higher plasma temperatures \citep{Guedel04.0},
this is qualitatively consistent with the shape of the thermal Bremsstrahlung continuum 
that increases towards shorter wavelengths for 
%$\log{T} > 5$; 
high temperatures \cite[see][]{Redfield02.0}. 
%Secondly, and probably more
%important are the relative contributions of continuum and line emission to the two GALEX
%bands. 
We recall here that the NUV band 
is dominated by chromospheric emission lines 
while the FUV band comprises many transition region emission lines.  
In this simplistic view, the observed dependence of 
${(FUV - NUV)}_{\rm exc}$ on X-ray flux indicates that higher
(X-ray) activity is related to a stronger enhancement of emission in the transition 
region with respect to the chromosphere. This is in agreement with the higher 
formation temperature of the FUV transition region lines ($T > 10^5$\,K) 
with respect to the NUV emission. 

We have examined flux-flux relations between H$\alpha$, NUV, FUV and X-ray emission.
The tighest correlation is observed for FUV vs NUV. As GALEX observes 
simultaneously in both bands, while the X-ray and H$\alpha$ observations are non-contemporaneous, 
this may indicate that the spread in the other relations is variability scatter.
 On the other hand, by comparing
the fluxes for stars with more than one measurement of 
a given activity diagnostic (UV or X-rays) we have demonstrated that long-term 
variability is not an important factor for the average activity level. Therefore, the
scatter in the flux-flux relations seems to represent an intrinsic range of physical conditions
between the chromosphere and corona of M dwarfs. 
While variability in general plays a minor role in this sample, in individual cases
it can be important. Comparing GALEX fluxes with published HST fluxes for two 
planet host stars we found one order of magnitude differences likely produced by
flaring. 

MD89 have studied the relation between Mg\,{\sc II} flux and X-ray flux for M
dwarfs observed with {\em IUE}, {\em EXOSAT} and {\em Einstein}. 
They found that the emission in these two observables
are correlated for dMe stars while they are not correlated for dM stars. We have separated
the $10$-pc sample into H$\alpha$ emitters (dMe) and stars with upper limit to the 
H$\alpha$ flux (dM) and examined the flux-flux relations between UV and X-ray emission. 
For the $10$-pc sample the trend towards larger scatter for the 
dM stars is confirmed but they are generally located on the extension
of the flux-flux relation of dMe stars. In particular, several dM stars do show
X-ray and/or chromospheric UV emission making the definition of dM as `non-active' 
versus dMe as `active' questionable. 
A similar conclusion was drawn by \cite{Walkowicz08.0} who found substantial UV 
and X-ray flux in a number of M dwarfs without H$\alpha$ emission. 

For X-rays vs. H$\alpha$ the power-law slope for the flux-flux relation of the $10$-pc sample is 
%consistent with the slope deteremined by \cite{MartinezArnaiz11.2} for .... and 
slightly steeper than the results of \cite{Stelzer12.0} for a sample of M dwarfs that 
comprised also late-M spectral types. 
The ultracool sample from \cite{Stelzer12.0} is likely biased towards strongly active
objects. However, \cite{Stelzer12.0} constructed their $F_{\rm surf,X}$ vs $F_{\rm surf,H\alpha}$
relation using the faintest measured flux level  in case of several epochs of
data for a given object. Therefore, their steeper slope may indicate a change of X-ray-to-H$\alpha$
flux ratio in the low-activity regime. The upper limits of the $10$-pc sample are not
sensitive enough to test this hypothesis. 

Despite the fact that we have exploited the most sensitive available observations for this
sample of nearby M dwarfs, a substantial number of objects ($30-40$\,\%) remain undetected in the
FUV and X-ray regime. The sensitivity limit may be responsible for the increase 
of the lower envelope of the values for the activity index in the X-ray band with
increasing $T_{\rm eff}$. The largest spread of the activity indices
is observed at spectral type $\sim$\,M4 where the range of rotation rates is largest. 
We have computed the average values of the activity indices for the subsample of 
fast rotators ($v \sin{i} > 5$\,km/s). 
While there are clear signs for saturation in H$\alpha$ and X-rays, the scatter of
the data points for fast rotators in the NUV and FUV range is large. 
Comparing the fractional flux output of X-rays, FUV, NUV and H$\alpha$ we find an increase
with energy for the four diagnostics. 
This trend was already noted by \cite{Hawley96.1} who found 
$\log{(L_{\rm x}/L_{\rm H\alpha})} \sim 0.5$ in a smaller volume-limited sample. 
For the $10$-pc sample 
the saturation levels for X-rays and H$\alpha$ differ by nearly one order of magnitude.

Finally, to study the age evolution of the activity in M dwarfs we have analysed the UV and
X-ray emission of a young comparison sample composed of the M stars from the TWA. These
stars are by about a factor hundred younger than the stars from the $10$-pc sample yet accretion
has stopped in the majority of them such that the observed emission can be attributed
to magnetic activity. We found that between $\sim 10$\,Myr and few\,Gyrs the X-ray, FUV 
and NUV luminosities drop by almost three orders of magnitude. A small group of young
M dwarfs in the $10$-pc sample bridges the gap between the luminosities of the TWA stars
and the bulk of the field M dwarfs indicating that the decrease of activity with age
is a continuous process. We have calculated the slopes of the age decay for the
X-ray, FUV and NUV luminosities. \cite{SanzForcada11.0} have calculated the age-luminosity
relation in the EUV band ($100-920$\,\AA) for a sample of planet host stars with
spectral types F-K. They found a
somewhat steeper slope than we do for the UV and X-rays bands bracketing the EUV. 
As the stars in our $10$-pc sample and possibly also 
those in the TWA have an unknown age spread that we did not take into account, we do not
give weight to the absolute values of our slopes. However, we can state that  
the age decay of the luminosity of M dwarfs is steeper for
the X-ray band than for the NUV and FUV. This is similar to analogous studies on solar analogs (G stars) \citep{Ribas05.1, Claire12.0}.

\section*{Acknowledgments}

BS would like to thank J.Sanz-Forcada and C.Cecchi-Pestellini 
for helpful conversations, and A.Reiners for providing the table with optical spectroscopic
measurements before its official availability. We appreciate the input of an anonymous referee. 
This work is based on archival data from the GALEX, XMM-Newton and ROSAT space missions. 
Partial financial support from ASI is acknowledged. 

\bibliographystyle{mn2e_fix} %mn2e.bst
\bibliography{mstarsMN}

%\clearpage

\begin{onecolumn}\begin{table*}\begin{minipage}{\linewidth}
\caption{Stellar parameters for $10$-pc M dwarf sample.\label{tab:tenpc_params_ste}}
\centering
\begin{tabular}{llccccc} \hline
Name  & Gl/GJ & $d$  & SpT & $T_{\rm eff}$ & $\log{f_{\rm bol, Earth}}$ & $v \sin{i}$ \\
      &       & [pc] &     & [K]           & [${\rm erg/cm^2/s}$]       & [km/s] \\ \hline
PM I00054-3721   & Gl  1          & $4.34 \pm 0.02$ & M1.5 & 3575 & $-7.37$ & $<2.5$ \\
PM I00115+5908   &  & $9.23 \pm 0.12$ & M5.5 & 2936 & $-9.33$ & ... \\
PM I00154-1608   & GJ 1005 AB        & $4.99 \pm 0.25$ & M4.0 & 3165 & $-8.44$ & $<3.0$ \\
PM I00184+4401   & Gl  15  B         & $3.56 \pm 0.89$ & M3.5 & 3241 & $-8.01$ & $<3.1$ \\
PM I01025+7140   & Gl  48            & $8.24 \pm 0.08$ & M3.0 & 3318 & $-8.58$ & $<2.5$ \\
PM I01026+6220   & Gl  49            & $9.96 \pm 0.15$ & M1.5 & 3575 & $-8.10$ & $<3.4$ \\
PM I01103-6726   & Gl  54            & $8.20 \pm 0.16$ & M2.0 & 3500 & $-8.09$ & ... \\
PM I01125-1659   & Gl  54.1          & $3.69 \pm 0.12$ & M4.5 & 3089 & $-8.30$ & $<2.5$ \\
\hline
\multicolumn{7}{l}{This table is available in its entirety in the electronic edition of MNRAS. Here only the first eight rows are} \\
\multicolumn{7}{l}{shown to illustrate the format.} \\
\end{tabular}\end{minipage}\end{table*}
\end{onecolumn}

\begin{onecolumn}\begin{table*}\begin{minipage}{\linewidth}
\caption{Observed NUV, FUV, and X-ray fluxes for $10$-pc sample. Flux units are ${\rm erg/cm^2/s}$. Lower and upper values for the fluxes derived from the uncertainties are given in brackets.\label{tab:tenpc_params_act}}
\normalsize\centering
\begin{tabular}{llcclcclcc} \hline
Name  &  & $\log{f_{NUV}}$ &  $\log{f_{NUV,lo},f_{NUV,up}}$ &  & $\log{f_{FUV}}$ &  $\log{f_{FUV,lo},f_{FUV,up}}$ &  & $\log{f_{X}}$ &  $\log{f_{X,lo},f_{X,up}}$ \\ \hline
PM I00054-3721   & $$ & $-13.14$ & $[-13.10,-13.17]$ & $<$ & $-13.76$ & $$ & $$ & $-13.01$ & $[-12.96,-13.07]$ \\
PM I00115+5908   & $$ & $$ & $$ & $$ & $$ & $$ & $<$ & $-13.20$ & $$ \\
PM I00154-1608   & $$ & $-14.15$ & $[-13.93,-14.37]$ & $<$ & $-13.71$ & $$ & $$ & $-13.41$ & $[-13.29,-13.58]$ \\
PM I00184+4401   & $$ & $$ & $$ & $$ & $$ & $$ & $$ & $-11.89$ & $[-11.85,-11.94]$ \\
PM I01025+7140   & $$ & $-12.97$ & $[-12.93,-13.02]$ & $$ & $-13.22$ & $[-12.95,-13.48]$ & $$ & $-13.19$ & $[-13.03,-13.43]$ \\
PM I01026+6220   & $$ & $$ & $$ & $$ & $$ & $$ & $<$ & $-13.27$ & $$ \\
PM I01103-6726   & $$ & $-12.97$ & $[-12.93,-13.01]$ & $$ & $-13.21$ & $[-13.00,-13.43]$ & $<$ & $-12.87$ & $$ \\
PM I01125-1659   & $$ & $-13.10$ & $[-13.05,-13.14]$ & $$ & $-12.70$ & $[-12.60,-12.80]$ & $$ & $-12.41$ & $[-12.36,-12.45]$ \\
%PM I00054-3721   & $$ & $-13.14$ & $$ & $<$ & $-13.76$ & $$ & $$ & $-13.01$ & $[-12.96,-13.07]$ \\
%PM I00115+5908   & $$ & $$ & $$ & $$ & $$ & $$ & $<$ & $-13.20$ & $$ \\
%PM I00154-1608   & $$ & $-14.15$ & $$ & $<$ & $-13.71$ & $$ & $$ & $-13.41$ & $[-13.29,-13.58]$ \\
%PM I00184+4401   & $$ & $$ & $$ & $$ & $$ & $$ & $$ & $-11.89$ & $[-11.85,-11.94]$ \\
%PM I01025+7140   & $$ & $-12.97$ & $[-12.93,-13.02]$ & $$ & $-13.22$ & $[-12.95,-13.48]$ & $$ & $-13.19$ & $[-13.03,-13.43]$ \\
%PM I01026+6220   & $$ & $$ & $$ & $$ & $$ & $$ & $<$ & $-13.27$ & $$ \\
%PM I01103-6726   & $$ & $-12.97$ & $[-12.93,-13.01]$ & $$ & $-13.21$ & $[-13.00,-13.43]$ & $<$ & $-12.87$ & $$ \\
%PM I01125-1659   & $$ & $-13.10$ & $[-13.05,-13.14]$ & $$ & $-12.70$ & $[-12.60,-12.80]$ & $$ & $-12.41$ & $[-12.36,-12.45]$ \\
\hline
\multicolumn{10}{l}{This table is available in its entirety in the electronic edition of MNRAS. Here only the first eight rows are shown to illustrate the format.} \\
\end{tabular}\end{minipage}\end{table*}
\end{onecolumn}

%\clearpage
%
%\include{tenpc_params_ste_MN_1}
%
%\addtocounter{table}{-1}
%
%\clearpage
%
%\include{tenpc_params_ste_MN_2}
%
%\addtocounter{table}{-1}
%
%\clearpage
%
%\include{tenpc_params_ste_MN_3}
%
%\addtocounter{table}{-1}
%
%\clearpage
%
%\include{tenpc_params_ste_MN_4}
%
%\clearpage
%
%\include{tenpc_params_act_MN_1}
%
%\addtocounter{table}{-1}
%
%\clearpage
%
%\include{tenpc_params_act_MN_2}
%
%\addtocounter{table}{-1}
%
%\clearpage
%
%\include{tenpc_params_act_MN_3}
%
%\addtocounter{table}{-1}
%
%\clearpage
%
%\include{tenpc_params_act_MN_4}

\label{lastpage}

\end{document}